\documentclass[12pt]{article}
\usepackage{amsmath}
\usepackage{graphicx}
\usepackage{enumerate}
\usepackage{natbib}
\usepackage{url} 
\usepackage{bm}
\usepackage{amsfonts}
\usepackage[flushleft]{threeparttable}
\usepackage{booktabs}
\usepackage{xcolor}
\usepackage[medium,compact]{titlesec}
\usepackage{tikz}
\usepackage{multirow}
\usepackage{subcaption}
\usepackage{bigints}
\usepackage{amsthm}
\usepackage{hyperref}
\usepackage{paralist}
\usepackage[nodisplayskipstretch]{setspace}
\usepackage{comment}
\usepackage{tikz}
\usetikzlibrary{decorations.pathreplacing,decorations.pathmorphing,arrows.meta,positioning,calc}

\allowdisplaybreaks

\newcommand{\blind}{1}

\newtheorem{remark}{Remark}

\newtheorem{assumption}{Assumption}
\newtheorem{theorem}{Theorem}
\newtheorem{proposition}{Proposition}
\newtheorem{corollary}{Corollary}
\newtheorem{definition}{Definition}
\theoremstyle{plain}

\newcommand{\bcx}{{\bm X}}

\newcommand{\bcy}{{\bm Y}}

\newcommand{\bco}{{\bm O}}

\newcommand{\bca}{{\bm A}}

\newcommand{\E}{\mathbb{E}}
\newcommand{\Prob}{\mathbb{P}}

\allowdisplaybreaks

\begin{document}

\def\spacingset#1{\renewcommand{\baselinestretch}%
{#1}\small\normalsize} \spacingset{1}


\if1\blind
{
  \title{\bf \Large Nonparametric efficient inference for network quantile causal effects under partial interference
  }
  \author{Chao Cheng$^{1,*}$ and Fan Li$^{2,\dagger}$\vspace{0.2cm}\\
  $^1$Department of Statistics and Data Science, \\
  Washington University in St. Louis\\
    $^2$Department of Biostatistics, Yale School of Public Health\\
    $^*$chaoc@wustl.edu \quad  $^\dagger$fan.f.li@yale.edu }
  \maketitle

  \vspace{-1cm} 
  
} \fi

\if0\blind
{
  \bigskip
  \bigskip
  \bigskip
  \begin{center}
    {\LARGE\bf Nonparametric efficient inference for network quantile causal effects under partial interference}
\end{center}
  \medskip
} \fi

\begin{abstract}
\noindent Interference arises when the treatment assigned to one individual affects the outcomes of other individuals. Commonly, individuals are naturally grouped into clusters, and interference occurs only among individuals within the same cluster, a setting referred to as partial interference. We study network causal effects on outcome quantiles in the presence of partial interference. We develop a general nonparametric efficiency theory for estimating these network quantile causal effects, which leads to a nonparametrically efficient estimator. The proposed estimator is consistent and asymptotically normal with parametric convergence rates, while allowing for flexible, data-adaptive estimation of complex nuisance functions. We leverage a three-way cross-fitting procedure that avoids direct estimation of the conditional outcome distribution. Simulations demonstrate adequate finite-sample performance of the proposed estimators, and we apply the methods to a clustered observational study. 
\end{abstract}

\noindent%

\spacingset{1.7} 

\section{Introduction}

\subsection{Background and motivating research question}

Quantile treatment effects represent a popular class of estimands in causal inference, particularly when outcome distributions are multi-modal, skewed, or heavy-tailed, rendering average treatment effects an inadequate summary of causal impacts. Within the potential outcomes framework, quantile treatment effects are defined as contrasts of population quantiles of the potential outcome distributions under alternative treatment assignments \citep{firpo2007efficient}, characterizing treatment impacts across different regions of the potential outcome distribution. 
A growing body of literature has studied quantile causal effects across a range of settings. These include quantile treatment effects with point or time-varying treatments under the ignorability assumption \citep[e.g.,][]{firpo2007efficient,cattaneo2010efficient,cheng2024doubly,kallus2024localized}; quantile treatment effects with instrumental variables \citep[e.g.,][]{abadie2002instrumental,chernozhukov2005iv,frolich2013unconditional}; quantile mediation effects in causal mediation analysis \citep{hsu2023doubly,cheng2025inverting}; and quantile treatment effects in regression discontinuity designs \citep{frandsen2012quantile}. 


Despite these developments, existing causal quantile methods critically relies on the Stable Unit Treatment Value Assumption (SUTVA), ruling out \textit{interference} or \textit{spillover effects} by assuming that an individual’s outcome does not depend on the treatment of other individuals. SUTVA can fail to hold in  applications where individuals are naturally connected with one another. 
Recently, a burgeoning literature has emerged to address interference across individuals \citep{hudgens2008toward,Aronow2017estimating,forastiere2021identification,ogburn2024causal}. A fundamental setting in this strand of literature is partial interference, in which individuals are grouped into disjoint clusters (e.g., villages), and interference only occurs among individuals within the same cluster \citep{sobel2006randomized}. Under partial interference, an individual’s outcome may be affected by their own treatment, corresponding to a direct effect, as well as by the treatments received by other individuals in the same cluster, corresponding to spillover effects. Sometimes, both spillover and direct effects are collectively referred to as network causal effects \citep{park2022efficient}. The goal of this article is to develop efficient nonparametric methods for evaluating a general class of network causal effects on outcome quantiles where partial interference is present.

Under partial interference, several prior efforts have examined network causal effects in two-stage cluster randomized trials under either finite- or super-population frameworks (e.g., \cite{sobel2006randomized,hudgens2008toward,liu2014large,basse2018analyzing,imai2021causal}). Others have considered clustered observational studies under ignorable assignments (e.g., \cite{tchetgen2012causal,liu2019doubly,papadogeorgou2019causal,park2024minimum,lee2025efficient}). 
Typically, network causal effects are defined as contrasts in average potential outcomes under different \textit{counterfactual treatment allocation policies}, which specify how treatment is assigned across individuals within a cluster. Examples include deterministic allocation policy \citep{qu2021semiparametric,hu2022average}, uniform allocation policy \citep{hudgens2008toward,tchetgen2012causal},  propensity score shift policies \citep{papadogeorgou2019causal,kennedy2019nonparametric,barkley2020causal,lee2025efficient},  and minimum resource threshold policy \citep{park2024minimum}. These policies provide interpretable cluster-level treatment allocation strategies and are well suited for settings with partial interference. 

Existing methods concerning interference  focus on causal effects on potential outcome means. One exception is \cite{shankar2024estimating}, which studies counterfactual distribution for randomized experiments under interference, but does not consider spillover causal effects as separate estimands. To our knowledge, there has been no general framework for evaluating network causal effects targeting potential outcome quantiles, which would enable a richer characterization of distributional impacts and treatment effect heterogeneity across the outcome distribution. For example, when studying the effect of smoking behavior on students' academic performance as measured by grade point average (GPA), the primary concern may not be the average effect, but rather the impact on students in the lower tail of the GPA distribution, where poor academic performance is most consequential. This is precisely the pursuit of the present article: we place quantile causal effects at the center of the analysis, developing a general framework for defining, identifying, and estimating causal effects under interference at arbitrary regions of outcome distribution.

\subsection{Our contributions}

We develop a semiparametric framework for quantile causal inference under partial interference. First, we define a collection of network quantile causal effects that capture both direct and spillover effects under a broad class of treatment allocation policies. 
Second, we establish a general nonparametric efficiency theory for the quantile estimands, leading to nonparametrically efficient estimators that achieve parametric convergence rates while accommodating flexible, data-adaptive estimation of nuisance parameters \citep{chernozhukov2018double}. A primary challenge in efficient quantile causal inference is the need to estimate the nuisance parameter corresponding to the entire conditional outcome distribution given baseline covariates \citep{cheng2025inverting}. To circumvent this challenge, we consider a three-way cross-fitting procedure \citep{kallus2024localized} that avoids direct estimation of the conditional outcome distribution. Instead, our approach requires only estimation of the conditional mean of a threshold outcome given covariates, thereby enabling the use of a broader class of machine learning methods for nuisance estimation in the causal quantile setting. Third, we propose a consistent and computationally efficient estimator of the asymptotic variance of the nonparametric estimator. Unlike the traditional variance estimator in quantile inference, our variance estimator requires no additional nuisance estimation beyond what is needed to construct the point estimator, making it suitable for large-scale applications. Collectively, our results provide a principled foundation for studying distributional causal effects under partial interference. 

{As an extension of existing propensity score shift policies, this paper also introduces a new cluster-level incremental propensity score policy (CPS), which is well suited for observational studies under partial interference. Existing propensity score shift policies are typically formulated either by assuming independence in treatment assignment across same-cluster individuals, as in the incremental propensity score policy of \cite{lee2025efficient}, or by relying on parametric models for the treatment assignment mechanism \citep{papadogeorgou2019causal,barkley2020causal}. In contrast, CPS directly perturbs the odds of the cluster propensity score, naturally accommodating within-cluster dependence in treatment assignment while avoiding reliance on a correctly specified parametric treatment model.}

As a motivating data example, we consider an application based on the National Longitudinal Study of Adolescent to Adult Health (Add Health) \citep{harris2019cohort}. 
Add Health collects rich information on students’ academic, behavioral, and demographic characteristics, as well as friendship networks. Leveraging these features, we study quantile effects of adolescents' smoking behavior on academic performance, measured by the total GPA across four core subjects (mathematics, English, history, and science). We construct a subcohort study from the Add Health data in which each student is grouped with their two closest friends based on friendship nominations, forming three-person clusters that naturally capture social interactions. This  allows us to examine both the direct effect of a student’s own smoking behavior and spillover effects arising from the smoking behavior of close peers, with a focus on network quantile causal effects beyond the average.

\section{Problem setup}\label{sec:data_structure}

\subsection{Data structure and notations}

Consider a study consisting of $n$ disjoint clusters. Let $M_i$ denote the number of individuals in cluster $i \in \{1,\dots,n\}$. For each individual $j \in \{1,\dots, M_i\}$ in cluster $i$, let $\bcx_{ij} \in \mathbb{R}^{d_{\bm X}}$ be a set of baseline covariates, $A_{ij} \in \{0,1\}$ be a binary treatment, $Y_{ij} \in \mathcal Y \subseteq \mathbb{R}$ be a continuous outcome of interest. Let $\bcx_i=[\bcx_{i1}^T,\dots, \bcx_{iM_i}^T]^T$, $\bca_i = [A_{i1},\dots,A_{iM_i}]^T$, and $\bcy_{i} = [Y_{i1},\dots,Y_{iM_i}]^T$ be the collection of covariates, treatment, and outcome for all individuals in cluster $i$. Conditioning on $M_i = m_i$, let $\mathcal X(m_i) \subseteq \mathbb{R}^{m_i\times d_{\bm X}}$, $\mathcal A(m_i) \subseteq \{0,1\}^{\otimes m_i}$, and $\mathcal Y(m_i) \subseteq \mathbb{R}^{m_i} $ be the support of $\bcx_i$, $\bca_i$, and $\bcy_i$, respectively. 

Under partial interference, an individual’s  outcome may depend on both the individual’s own treatment and the treatment assignments of other individuals within the same cluster. Under the potential outcomes framework, let $Y_{ij}(\bm a_i)$ denote the potential outcome of individual $j$ in cluster $i$ that would be observed if the treatment vector for cluster $i$ were set to $\bca_i = \bm a_i \in \mathcal{A}(M_i)$. 
Sometimes, we also decompose the cluster treatment vector as $\bca_i = (A_{ij}, \bca_{i(-j)})$, where $\bca_{i(-j)} = (A_{i1}, \ldots, A_{i(j-1)}, A_{i(j+1)}, \ldots, A_{iM_i})$ is the cluster treatment vector excluding individual $j$, and correspondingly write $\bm a_i = (a_{ij}, \bm a_{i(-j)})$. With this notation, the potential outcome can be equivalently expressed as $Y_{ij}(\bm a_i) = Y_{ij}(a_{ij}, \bm a_{i(-j)})$; this notation explicitly distinguishes an individual’s own treatment ($a_{ij}$) from the treatments of the remaining cluster members ($\bm a_{i (-j)}$). Defining $\bm Y_i (\bm a_i) = [Y_{i1}(\bm a_i),\dots, Y_{iM_i}(\bm a_i)]^T$, we then write the collection of all, even though not fully observed, random variables in a cluster as $\mathcal W_i = \{M_i,\bm X_i, \bm A_i, \bm Y_i (\bullet)\}$, where $\bm Y_i(\bullet) = \{\bm Y_i(\bm a_i), \bm a_i \in \mathcal A(M_i)\}$. We adopt the following super-population framework to characterize the distribution of $\mathcal W_i$:
\begin{assumption}\label{assum:super_population}
(Super-population)  $\{\mathcal W_1,\dots,\mathcal W_n\}$ are mutually independent. The cluster size $M_i$ follows an unknown distribution $\mathbb{P}_{M_i}$ with finite support $\mathbb{M} \subset \mathbb{N}^{+}$. Conditioning on $M_i$, the joint distribution of $\{\bm X_i, \bm A_i, \bm Y_i (\bullet)\}$ has the decomposition $\mathbb{P}_{\bm Y_i (\bullet),\bm A_i,\bm X_i|M_i} = \mathbb{P}_{\bcy_i(\bullet)|\bca_i,\bcx_i,M_i}\times \mathbb{P}_{\bca_i|\bcx_i,M_i}\times \mathbb{P}_{\bcx_i|M_i}$ with all components having a finite second moment. 
\end{assumption}

Under Assumption \ref{assum:super_population}, the observed data in a cluster, $\bco_i = \{M_i,\bcx_i,\bca_i,\bcy_i\}$, are independently and identically distributed across clusters. The observed data distribution can be decomposed as $\mathbb{P}_{\bm O_i}=\mathbb{P}_{\bcy_i|\bca_i,\bcx_i,M_i}\times \mathbb{P}_{\bca_i|\bcx_i,M_i}\times \mathbb{P}_{\bcx_i|M_i}\times \mathbb{P}_{M_i}$.

\subsection{Network quantile causal effects}


Under partial interference, network causal effects are defined with respect to treatment allocation policies \citep{hudgens2008toward,diaz2011population,lee2025efficient} that specify how treatments are jointly assigned for all individuals within a cluster. Given a cluster of size $M_i$ with covariates $\bm X_i$, we consider a hypothetical stochastic intervention that assigns the cluster treatment vector to $\bm A_i = \bm a_i \in \mathcal{A}(M_i)$ with probability $H(\bm a_i | \bm X_i, M_i)$, where $H \equiv H(\bullet | \bm X_i, M_i)$ is a user-specified probability mass function representing a treatment allocation policy.  This paper treats $H(\bm a_i | \bm X_i, M_i)$ as either known or as a function of the cluster propensity score $\pi(\bm a_i | \bm X_i, M_i) := \mathbb{P}(\bm A_i = \bm a_i | \bm X_i, M_i)$, where the latter characterizes the observed treatment assignment mechanism under the data-generating process described in Assumption~\ref{assum:super_population}. Different choices of $H$ encode different treatment allocation policies, where example policies with their meanings will be introduced in Section \ref{sec:policies}.


Given policy $H$, we define the realized cluster treatment vector as $\bm A_i^H$, which is a random draw from $\mathcal{A}(M_i)$ according to $H(\bullet | \bm X_i, M_i)$. The corresponding policy-specific potential outcome for individual $j$ in cluster $i$ is $Y_{ij}(\bm A_i^H)$, representing individual $j$'s outcome that would be observed if the cluster received treatment policy $\bm A_i^H$. By construction, $Y_{ij}(\bm A_i^H) = Y_{ij}(\bm a_i)$ with probability $H(\bm a_i | \bm X_i, M_i)$. To formally define the network quantile causal effects, we envision the potential outcome of a randomly sampled individual within a given cluster. Specifically, let $J_i$ be a random draw based on a uniform distribution on $\{1,\dots,M_i\}$, which encodes a random individual in a given cluster. The policy-specific potential outcome for this random individual is $Y_{iJ_i}(\bm A_i^H)$, which can be equivalently written as  $Y_{iJ_i}(\bm A_i^H) \equiv Y_{iJ_i}(A_{iJ_i}^H,\bm A_{i(-J_i)}^H)$. The distribution of $Y_{iJ_i}(\bm A_i^H)$, induced by policy $H$ and the random individual index $J_i$, leads to the following definition of our causal estimand.
\begin{definition}[Quantiles of the policy-specific potential outcome]\label{def:quantile}
For $q\in(0,1)$, the marginal $q$-th quantile of the policy-specific potential outcome is defined as
$$
Q_H^{(\star)}(q) = \inf \left\{\theta \in \mathbb{R}: \mathbb{P}\!\left(Y_{iJ_i}(\bm A_i^H)\leq \theta \right) \geq q \right\},
$$
where $J_i \sim \text{Uniform}\{1,\dots,M_i\}$, $\bm A_i^H \sim H(\bullet | \bm X_i, M_i)$, and $J_i \perp \bm A_i^H \mid (\bm X_i,M_i)$. To facilitate the characterization of direct and spillover effects, for $a\in\{0,1\}$, we further define
$$
Q_H^{(a)}(q) = \inf\left\{\theta \in \mathbb{R}:\mathbb{P}\!\left(Y_{iJ_i}(a,\bm A_{i(-J_i)}^H) \leq \theta\right) \geq q\right\}.
$$
\end{definition}
By construction, $Q_H^{(\star)}(q)$ represents the $q$-th quantile of the potential outcome distribution for a randomly selected individual when cluster treatments are assigned according to policy $H$.  In contrast, for $a\in\{0,1\}$, $Q_H^{(a)}(q)$ denotes the $q$-th quantile of the potential outcome for that same randomly selected individual when their own treatment is fixed to $a$, while the treatments of others in the cluster continue to follow policy $H$. Thus, $Q_H^{(a)}(q)$ separates the effect of individual treatment from the influence of peer treatment assignments, thereby enabling the definition of direct and spillover effect on the scale of quantiles.

Finally, we introduce a set of network quantile causal effects based on pair contrasts among $Q_H^{(\star)}(q)$, $Q_H^{(0)}(q)$, and $Q_H^{(1)}(q)$. For two treatment allocation policies $H$ and $H'$, we first define the \textit{overall quantile effect} (OQE) as $\text{OQE}(q; H, H') := Q_H^{(\star)}(q) - Q_{H'}^{(\star)}(q)$, which compares how changing the treatment allocation policy among all cluster members shifts the $q$-th quantile of the potential outcome distribution. To study the direct effect of changing an individual's own treatment, we define the \textit{direct quantile effect} (DQE) as $\text{DQE}(q; H) :=Q_H^{(1)}(q) - Q_H^{(0)}(q)$, which represents the causal contrast between treating and not treating an individual, while holding the treatments from other same-cluster members to policy $H$. Similarly, to measure the treatment effect due to interference, we define the \textit{spillover quantile effect} as $\text{SQE}_a(q; H,H') :=Q_H^{(a)}(q) - Q_{H'}^{(a)}(q)$, with $a\in\{0,1\}$, which compares the potential outcomes for an individual with fixed treatment $a$ while changing the policy of other cluster members from $H'$ to $H$. At the end, we introduce the \textit{total quantile effect} (TQE) as  $\text{TQE}(q; H, H') := Q_H^{(1)}(q) - Q_{H'}^{(0)}(q)$, which captures the combined influence of changing both the individual’s own treatment status and the treatment allocation policy among other same-cluster members. It is straightforward to verify $\text{TQE}(q; H, H') = \text{SQE}_1(q; H,H') + \text{DQE}(q; H') = \text{SQE}_0(q; H,H') + \text{DQE}(q; H)$. 

\begin{remark}
(Alternative definition of quantile estimands) The quantile estimands in Definition \ref{def:quantile} are defined as the (generalized) inverse of the cumulative distribution functions (CDFs)
$\mathbb{P}\!\left(Y_{iJ_i}(\bm A_i^H) \leq \theta\right)$ and 
$\mathbb{P}\!\left(Y_{iJ_i}(a,\bm A_{i(-J_i)}^H) \leq \theta\right)$.
We show in Appendix B.1 that these CDFs admit the following alternative representations:  
\begin{align*} 
\mathbb{P}\left(Y_{iJ_i}(\bm A_i^H)\leq \theta \right) & = \E\left[\frac{1}{M_i}\sum_{i=1}^{M_i}\sum_{\bm a_i\in \mathcal A(M_i)} H(\bm a_i \mid \bm X_i, M_i) \times \mathbb{I}(Y_{ij}(\bm a_i) \leq \theta)\right], \\
\mathbb{P}\!\left(Y_{iJ_i}(a,\bm A_{i(-J_i)}^H) \leq \theta\right) & = \E\left[\frac{1}{M_i}\sum_{i=1}^{M_i}\sum_{\bm a_{i(-j)}\in \mathcal A(M_i-1)} H(\bm a_{i(-j)} \mid \bm X_i, M_i) \times \mathbb{I}(Y_{ij}(a,\bm a_{i(-j)}) \leq \theta)\right],
\end{align*}
where $H(\bm a_{i(-j)}|\bm X_i, M_i) = H(1,\bm a_{i(-j)}|\bm X_i, M_i)+H(0,\bm a_{i(-j)}|\bm X_i, M_i)$. These alternative representations clarify that the policy-specific quantile estimands are nonlinear transformations of the same policy-specific estimands studied in the literature on network average causal effects (e.g., \cite{park2022efficient,lee2025efficient}). In particular, if one replaces the threshold-transformed potential outcomes $\mathbb{I}\!\{Y_{ij}(\bullet)\leq\theta\}$ with the original potential outcomes $Y_{ij}(\bullet)$, the resulting expressions directly correspond to policy-specific average potential outcomes used to define network average causal effects. Hence, the quantile estimands naturally extend network average causal effects from mean functionals to distributional functionals.
\end{remark}

\subsection{Example treatment allocation policies}\label{sec:policies}

We focus on studying the following four treatment allocation policies, although our theory applies to general choices of policies. Each policy is expressed by $H_{\texttt{par}}^{\texttt{name}}(\bm a_i | \bm X_i,M_i)$, where the superscript ``\texttt{name}" denotes the specific policy and the subscript``\texttt{par}" represents any hyperparameter associated with the policy.

{1. \emph{Deterministic allocation policy (DAP).}}  
Indexed by $a\in\{0,1\}$, the  DAP is defined by $H_a^{\texttt{DAP}}(\bm a_i | \bm X_i,M_i)=1$ if $\bm a_i = a \times \bm{1}_{M_i}$ but $H_a^{\texttt{DAP}}(\bm a_i | \bm X_i,M_i)=0$ for all other values of $\bm a_i$, where $\bm{1}_{M_i}$ is a vector of 1 of length $M_i$. The policy represents extreme allocation strategies, shifting treatment from complete abstention ($a=0$) to full adoption ($a=1$). 

{2. \emph{Uniform allocation policy (UAP).}}  
Each individual is independently assigned treatment with probability $\alpha\in[0,1]$, yielding the allocation policy $H_\alpha^{\texttt{UAP}}(\bm a_i | \bm X_i,M_i)=\prod_{j=1}^{M_i}\alpha^{a_{ij}}(1-\alpha)^{1-a_{ij}}$ \citep{tchetgen2012causal}. This policy generalizes the DAP by allowing individuals to receive treatment with a specified probability. 

{DAP and UAP are useful in randomized trials, but they may be relatively less suitable for observational studies, where individuals may differ substantially in their baseline probabilities of receiving treatment. As a result, policies that assign treatment deterministically or uniformly within a cluster may be less realistic and may place substantial weight on treatment patterns that are poorly supported by the observed data. This motivates the next two policies, which generate more realistic counterfactual interventions by shifting the observed treatment assignment mechanism.}

{3. \emph{Individual-level incremental propensity score policy (IPS).}} 
Let $\pi_{ij}:=\mathbb{P}(A_{ij}=1 \mid \bm X_i,M_i)$ denote the individual propensity score. Indexed by $\delta>0$, the IPS is defined as
$H_{\delta}^{\texttt{IPS}}(\bm a_i\mid \bm X_i, M_i) = \prod_{j=1}^{M_i} (\pi_{ij}^\delta)^{a_{ij}}(1-\pi_{ij}^\delta)^{1-a_{ij}}$ \citep{lee2025efficient}, where $\pi_{ij}^\delta := \mathbb{P}_\delta(A_i=1 \mid \bm X_i,M_i ) = \delta\pi_{ij}/\{1-\pi_{ij}+\delta\pi_{ij}\}$ is the shifted individual propensity score under the policy \citep{kennedy2019nonparametric}. It is straightforward to show that 
\begin{equation}\label{eq:odds_ips}
\delta = \frac{\mathbb{P}_\delta(A_{ij}=1 \mid \bm X_i,M_i )}{\mathbb{P}_\delta(A_{ij}=0 \mid \bm X_i,M_i )} \Bigg/ \frac{\mathbb{P}(A_{ij}=1 \mid \bm X_i,M_i )}{\mathbb{P}(A_{ij}=0 \mid \bm X_i,M_i )},
\end{equation}
therefore the IPS modifies the individual propensity score distribution such that the counterfactual odds of individual treatment ${\mathbb{P}_\delta(A_{ij}=1 \mid \bm X_i,M_i )}/{\mathbb{P}_\delta(A_{ij}=0 \mid \bm X_i,M_i )}$ is $\delta$ times the factual odds of the individual treatment ${\mathbb{P}(A_{ij}=1 \mid \bm X_i,M_i )}/{\mathbb{P}(A_{ij}=0 \mid \bm X_i,M_i)}$. In other words, the IPS shifts the odds of individual treatment uniformly across individuals by a factor $\delta$, making treatment more likely when $\delta>1$ and less likely when $0<\delta<1$. 

A limitation of IPS is that it operates independently across individuals within a cluster. As a result, the IPS is more interpretable when the factual treatment mechanism is believed to be individually independent with $\pi(\bm a_i|\bm X_i,M_i) = \prod_{i=1}^{M_i} \mathbb{P}(A_{ij}=a_{ij} \mid \bm X_i,M_i)$. In many clustered settings, treatment decisions may be correlated due to shared contextual factors among individuals. In such cases, modifying individual propensities independently fails to preserve the structural dependence observed in the data \citep{papadogeorgou2019causal}. As a generalization of IPS, we further propose the following new policy preserving the structural dependence in the counterfactual cluster propensity score.

{4. \emph{Cluster-level incremental propensity score policy (CPS).}} Instead of shifting the individual propensity score, we can shift the cluster propensity score according to the total number of treated individuals in the cluster. Let $s(\bm a_i)=\sum_{j=1}^{M_i} a_{ij}$ denote total number of treated individuals under assignment $\bm a_i$. Indexed by $\delta>0$, we define the CPS as
$$
H_\delta^{\texttt{CPS}}(\bm a_i| X_i,M_i) =
\frac{\delta^{\,s(\bm a_i)}\,\pi(\bm a_i \mid \bm X_i,M_i)}
{\sum_{\bm a_i'\in\mathcal A(M_i)}\delta^{\,s(\bm a_i')}\,\pi(\bm a_i'\mid \bm X_i,M_i)}.
$$
The CPS induces a constant odds shift for the cluster treatment assignments. Similar to \eqref{eq:odds_ips}, one can show that, for any two assignments $\bm a_i \neq \bm a_i'$ with $s(\bm a_i)-s(\bm a_i')=1$, we have
\begin{equation*}
\delta = \frac{H_\delta^{\texttt{CPS}}(\bm a_i| \bm X_i,M_i)}{H_\delta^{\texttt{CPS}}(\bm a_i'| \bm X_i,M_i)}\Big/ \frac{\pi(\bm a_i| \bm X_i,M_i)}{\pi(\bm a_i'| \bm X_i,M_i)},
\end{equation*}
therefore the CPS modifies the cluster propensity score distribution such that the counterfactual odds of increasing one treated unit in cluster, ${H_\delta^{\texttt{CPS}}(\bm a_i| \bm X_i,M_i)}/{H_\delta^{\texttt{CPS}}(\bm a_i'| \bm X_i,M_i)}$, is $\delta$ times the factual odds of increasing one treated unit in cluster, ${\pi(\bm a_i| \bm X_i,M_i)}/{\pi(\bm a_i'| \bm X_i,M_i)}$.
In other words, treatment assignments with more treated individuals are upweighted (or underweighted) relative to the factual assignment when $\delta>1$ (and $0<\delta<1$); the case $\delta=1$ recovers the factual assignment mechanism. In the Add Health study, $\delta = 2$ (or $\delta = 0.5$) represents a counterfactual scenario in which the odds that a cluster contains one more smoker under the policy are doubled (or halved) relative to the observed treatment assignment mechanism. Under conditional independence of individual treatments within clusters (i.e., $\pi(\bm a_i|\bm X_i,M_i) = \prod_{i=1}^{M_i} \mathbb{P}(A_{ij}=a_{ij} \mid \bm X_i,M_i)$), the CPS reduces to the IPS. 


For all policies considered above, $H(\bm a_i|\bm X_i, M_i)$ is either known (for $H_{a}^{\texttt{DAP}}$ and $H_{\alpha}^{\texttt{UAP}}$) or depends on the true cluster propensity score (for $H_{\delta}^{\texttt{IPS}}$ and $H_{\delta}^{\texttt{CPS}}$).



\section{Identification and weighting-based estimation}\label{sec:identification}

\subsection{Assumptions and nonparametric identification}

The following assumptions are used to identify the network quantile causal effects with a given quantile in the range $q \in \mathcal Q := [\epsilon, 1-\epsilon]$, where $\epsilon$ is a fixed number within $(0,0.5)$. 

\begin{assumption}\label{assum:consist}
(Consistency) For all $\bm a_{i} \in \mathcal A(M_i)$, $Y_{ij}(\bm a_{i}) = Y_{ij}$ if $\bm A_i = \bm a_i$. 
\end{assumption}

\begin{assumption}\label{assum:ignor}
(Treatment ignorability)  $Y_{ij}(\bm a_{i}) \perp \bm A_i \mid \{\bm X_i, M_i\}$ for all $\bm a_{i} \in \mathcal A(M_i)$. 
\end{assumption}

\begin{assumption}\label{assum:positivity}
(Overlap)  Assume $\pi (\bm a_i|\bm X_i, M_i)>0$ for all $\bm a_{i} \in \mathcal A(M_i)$. Moreover, assume $\sum_{\bm a_i \in \mathcal A(M_i)} H(\bm a_i|\bm X_i, M_i) = 1$ such that the policy does not assign positive probability to treatment vectors outside $\mathcal A(M_i)$.
\end{assumption}

\begin{assumption}\label{assum:unique}
(Uniqueness of quantiles) There exists $\Theta_{\star} := (c_1^{(\star)}, c_2^{(\star)}) \in \mathbb{R}$ such that $\mathbb{P}\!\left(Y_{iJ_i}(\bm A_i^H)\leq \theta\right)$ is continuous and strictly increasing in  $\theta \in \Theta_{\star}$ with $\mathbb{P}\!\left(Y_{iJ_i}(\bm A_i^H)\leq c_1^{(\star)}\right) < \epsilon$ and $\mathbb{P}\!\left(Y_{iJ_i}(\bm A_i^H)\leq c_2^{(\star)}\right)  > 1- \epsilon$. For both $a\in\{0,1\}$, there exists $\Theta_{a} := (c_1^{(a)}, c_2^{(a)}) \in \mathbb{R}$ such that $\mathbb{P}\!\left(Y_{iJ_i}(a,\bm A_{i(-J_i)}^H)\leq \theta\right)$ is continuous and strictly increasing in  $\theta \in \Theta_a$ with $\mathbb{P}\!\left(Y_{iJ_i}(a,\bm A_{i(-J_i)}^H)\leq c_1^{(a)}\right) < \epsilon$ and $\mathbb{P}\!\left(Y_{iJ_i}(a,\bm A_{i(-J_i)}^H)\leq c_2^{(a)}\right) > 1- \epsilon$.
\end{assumption}

Assumption~\ref{assum:consist} requires that treatment is well defined and that no interference occurs across clusters. Assumption \ref{assum:ignor} states that, conditional on baseline covariates and cluster size, the treatment allocation is independent of the potential outcomes, which holds in randomized trials and is also plausible in clustered observational study when the measured confounders are sufficiently rich that the confounding between treatment and outcome is adequately controlled. Assumption \ref{assum:positivity} requires that the cluster individuals have a positive probability to receive treatment in the support $\mathcal A(M_i)$; it also assumes that support of the treatment allocation policy $H$ is no larger than support of true treatment assignment mechanism. Assumption~\ref{assum:unique} is a technical condition to guarantee that the relevant quantiles of the potential outcome distribution are uniquely defined, excluding flat regions in the distribution function over the quantile range of interest; this is a mild regularity condition and typically satisfied when the outcome is continuous \citep{firpo2007efficient}.

Throughout, we focus on identification and estimation of the counterfactual quantile estimands, $\{Q_{H}^{(\star)}(q), Q_{H}^{(1)}(q), Q_{H}^{(0)}(q)\}$, based on which all network quantile causal effects are  obtained because they are simple contrasts between the quantile estimands. Theorem shows that Assumptions \ref{assum:super_population}–\ref{assum:unique} are sufficient for identifying the three quantile estimands.

\begin{theorem}\label{thm: identification}
(Nonparametric identification of the policy-specific quantile estimands) Suppose that Assumptions \ref{assum:super_population}--\ref{assum:unique} hold. For any $t\in\{\star,0,1\}$ and $q\in \mathcal Q$, $Q_{H}^{(t)}(q)$ is the unique solution of the following population moment condition with respect to $\theta \in \Theta_t$:
$$
\E\left[\frac{1}{M_i}\sum_{j=1}^{M_i} \sum_{\bm a_i \in \mathcal A(M_i)} \frac{\mathbb{I}(\bm A_i = \bm a_i) w_{ij}^{(t)}(\bm a_i,\bm X_i, M_i)  \left\{\mathbb{I}\left(Y_{ij}\leq \theta\right) - q\right\} }{\pi(\bm a_i|\bm X_i, M_i)} \right] = 0,
$$
where $w_{ij}^{(\star)}(\bm a_i,\bm X_i, M_i) = H(\bm a_i|\bm X_i, M_i)$, $w_{ij}^{(a)}(\bm a_i,\bm X_i, M_i) = \mathbb{I}(a_{ij}=a) \times H(\bm a_{i(-j)}|\bm X_i, M_i)$ for $a\in\{0,1\}$, and $H(\bm a_{i(-j)}|\bm X_i, M_i) = H(1,\bm a_{i(-j)}|\bm X_i, M_i)+H(0,\bm a_{i(-j)}|\bm X_i, M_i)$. 
\end{theorem}

\subsection{Modeling the cluster propensity score}\label{sec:cluster_propensity_score}

According to Theorem \ref{thm: identification}, $\{\widehat \pi(\bm a_i|\bm X_i, M_i),\widehat w_{ij}^{(t)}(\bm a_i,\bm X_i, M_i)\}$ need to be calculated in order to infer the network quantile causal effect. In particular, we only need to estimate $\pi(\bm a_i|\bm X_i, M_i)$, as after obtaining $\widehat \pi(\bm a_i|\bm X_i, M_i)$, $\widehat w_{ij}^{(t)}(\bm a_i,\bm X_i,M_i)$ can be immediately calculated. This is because $w_{ij}^{(t)}(\bm a_i,\bm X_i,M_i)$ depends on $H(\bm a_i | \bm X_i,M_i)$, and the latter is either known (for DAP and UAP) or a function of $\pi(\bm a_i|\bm M_i, M_i)$ (for IPS and CPS). Consequently, computation of $\widehat w_{ij}^{(t)}(\bm a_i,\bm X_i,M_i)$ is either explicit or at most relies on  $\widehat{\pi}(\bm a_i | \bm X_i,M_i)$.

To proceed, we consider a hybrid copula approach to estimate $\pi(\bm a_i|\bm X_i, M_i)$, and several alternative modeling strategies are given in Web Appendix A.1. This approach allows flexible machine-learning methods for modeling the individual propensity score, while introducing a parsimonious copula to account for dependence across individuals, yielding a method that is more robust than fully parametric mixed-effects models and more scalable than fully nonparametric modeling of the cluster propensity score \citep{cheng2024semiparametric}. Let $\Pi_{ij}(a)=\mathbb{P}(A_{ij}\le a \mid \bm X_i,M_i)$ denote the CDF of $A_{ij}$ given $\{\bm X_i,M_i\}$. We assume that the joint CDF of $\bm A_i$ given $\{\bm X_i,M_i\}$ admits the following copula representation
\begin{equation}\label{eq:copula}
\mathbb{P}(A_{i1}\le a_{i1},\dots,A_{iM_i}\le a_{iM_i}\mid \bm X_i,M_i)
= \mathcal C_{\bm \gamma}\big(\Pi_{i1}(a_{i1}),\dots,\Pi_{iM_i}(a_{iM_i})\big),
\end{equation}
where the copula function $\mathcal C_{\bm \gamma}$ links the joint distribution of $(A_{i1},\dots,A_{iM_i})$ to their marginal distributions, and $\bm \gamma$ is a finite-dimensional parameter indexing the association among $\{A_{i1},\dots,A_{iM_i}\}$ conditional on $\{\bm X_i,M_i\}$. The cluster propensity score under this copula representation is
$
\pi(\bm a_i\mid \bm X_i,M_i)
= \sum_{\bm z \in \{0,1\}^{\otimes M_i}}(-1)^{\sum_{j=1}^{M_i} z_j}
\,\mathcal C_{\bm \gamma}\Big(\Pi_{i1}(a_{i1}-z_1),\dots,\Pi_{iM_i}(a_{iM_i}-z_{M_i})\Big),
$
where the summation is taken over all corners of the hypercube $\{0,1\}^{\otimes M_i}$. To fit this model, we first estimate the individual propensity scores $\pi_{ij}$ using any data-adaptive machine-learning methods for binary outcomes, yielding estimated marginal CDFs $\widehat \Pi_{ij}$. Next, the copula parameter $\bm\gamma$ is estimated via maximum pseudo-likelihood based on the implied copula model $\mathcal C_{\bm \gamma}\big(\widehat\Pi_{i1}(A_{i1}),\dots,\widehat \Pi_{iM_i}(A_{iM_i})\big)$. Finally, we plug $(\widehat \Pi_{i1},\dots, \widehat \Pi_{iM_i}, \widehat{\bm \gamma})$ into the expression for $\pi(\bm a_i | \bm X_i,M_i)$ to obtain the estimated cluster propensity score. The detailed procedures for performing this approach is given in Web Appendix A.1, which also gives a concrete example based on the Gaussian copula model for $\mathcal C_{\bm \gamma}$.

\subsection{Inverse probability weighting estimator}\label{sec:IPW}


We propose an inverse probability weighting (IPW) estimator that is directly motivated from the identification formula in Theorem \ref{thm: identification}. Since we consider using data-adaptive methods for estimating cluster propensity score, it is well known that direct plug-in IPW estimation may suffer from overfitting bias arising from using the same data to estimate propensity score and evaluate the target estimating equation \citep{chernozhukov2018double}. Therefore, we adopt the cross-fitting procedure to mitigate the overfitting bias.

To proceed, we take a $L\geq 2$ fold random partition $\{\mathcal D_1,\dots,\mathcal D_L\}$ of data indices $\{1,\dots,n\}$ such that each fold has approximately the same size. For any $t\in\{\star,0,1\}$, the cross-fitted IPW estimator of $Q_{H}^{(t)}(q)$, denoted by $\widehat{Q}_{H}^{(t),\texttt{ipw}}(q)$, is an  solution in $\theta$ to
\begin{equation}\label{eq:ipw}
\frac{1}{n}\sum_{l=1}^{L}\sum_{i\in \mathcal D_l}\Bigg[\frac{1}{M_i}\sum_{j=1}^{M_i}\sum_{\bm a_i\in\mathcal{A}(M_i)} 
\frac{\mathbb{I}(\bm A_i = \bm a_i) \widehat w_{ij,l}^{(t)}(\bm a_i,\bm X_i,M_i)\{\mathbb{I}(Y_{ij}\leq\theta)-q\}}
     {\widehat{\pi}_l(\bm a_i | \bm X_i,M_i)}
\Bigg] = 0,
\end{equation}
where $\{\widehat w_{ij,l}^{(t)},\widehat \pi_l\}$ are estimators of $\{w_{ij}^{(t)},\pi\}$ constructed by the machine-learning model in Section \ref{sec:cluster_propensity_score} that are trained based on the full data except $\mathcal D_l$.
The estimating equation above is the empirical analogue of the identification moment condition in Theorem \ref{thm: identification}, where the population expectation ``$\mathbb E$" is replaced by the cross-fitted sample average ``$\frac{1}{n}\sum_{l=1}^{L}\sum_{i\in \mathcal D_l}$". Define $\|\bullet\|_k$ as the $L_k(\mathbb{P})$-norm such that $\|f(\bco_i)\|_k = \left\{\int f^k(\bm o_i)d\mathbb{P}(\bm o_i)\right\}^{1/k}$, and let $O_{\mathbb{P}}$ and $o_{\mathbb{P}}$ be the big-$O$ and little-$o$ quantities with respect to $\mathbb{P}$. Then, the large-sample result for the IPW estimator is described in Proposition \ref{thm:ipw}.

\begin{proposition}\label{thm:ipw} 
For any $l\in\{1,\dots,L\}$,  suppose that (i) $0<\widehat\pi_{l}(\bm a_i| \bm x_i, m_i)<1$; (ii) the convergence rate of $\widehat\pi_{l}(\bm a_i, \bcx_i, M_i)$ satisfies
$\left\|\sum_{\bm a_i \in \mathcal A(M_i)}|(\widehat\pi_{l} - \pi)(\bm a_i, \bcx_i, M_i)|\right\|_{2} = O_\Prob(r_\pi)$; (iii) $\widehat{Q}_{H}^{(t),\texttt{ipw}}(q)$ is an approximate solution to \eqref{eq:ipw} with an error of order $o_\Prob(n^{-1/2})$. Then, for any policy in Section \ref{sec:policies}, we have that $\sup_{q\in \mathcal Q} | \widehat Q_H^{(t),\texttt{ipw}}(q) - Q_H^{(t)}(q) | = O_\Prob(\max\{r_\pi,n^{-1/2}\})$. 
\end{proposition}

Proposition \ref{thm:ipw} shows that the convergence rate of the IPW estimator is only guaranteed to be no slower than $O(\max\{r_\pi,n^{-1/2}\})$, which is generally slower than $O(n^{-1/2})$ as data-adaptive machine-learning models are used to estimate propensity scores. Therefore, the IPW estimator may be neither efficient nor $\sqrt{n}$-consistent. These limitations motivate the development of a nonparametric efficient estimator in the following section.

\section{Nonparametric efficient estimation}\label{sec:nonparametric_estimation}

\subsection{Efficient influence functions}

We next develop a nonparametric efficient estimator for $Q_{H}^{(t)}(q)$. Our construction is based on the efficient influence function (EIF) of $Q_{H}^{(t)}(q)$ under the nonparametric model of observed data distribution $\mathbb{P}:=\mathbb{P}_{\bm O_i}$. Based on the nonparametric causal inference theory \citep{chernozhukov2018double}, the EIF provides an estimand-centered framework enabling the construction of efficient estimators that achieve parametric-rate inference even when the nuisance functions are estimated at slower, nonparametric rates. Following \cite{bickel1993efficient}, to characterize the EIF, we define a parametric submodel $\mathbb{P}_{v}$ indexed by a one-dimensional parameter $v\in[0,1]$, where $\mathbb{P}_0=\mathbb{P}$ is the true observed data distribution. Denote the value of $Q_{H}^{(t)}(q)$ evaluated under $\mathbb{P}_{v}$ as $Q_{H}^{(t)}(q;\mathbb{P}_v)$. Notice that the true value of $Q_{H}^{(t)}(q)$ is $Q_{H}^{(t)}(q;\mathbb{P}_0)$, which is characterized by the estimating function in Theorem \ref{thm: identification}. The nonparametric EIF of $Q_{H}^{(t)}(q)$, whenever exists, is the unique, zero-mean finite-variance function $\psi_H^{(t)}\left(\bm O_i\right)$ satisfying $\frac{d}{d v} Q_{H}^{(t)}(q;\mathbb{P}_v)\mid_{v=0} = \E[\psi_H^{(t)}\left(\bm O_i\right)  \mathcal S(\bm O_i)]$, where $\mathcal S(\bm O_i)$ is the likelihood score of $\mathbb{P}_{v}$.

To derive $\psi_H^{(t)}\left(\bm O_i\right)$, we invoke a regularity condition regarding the treatment allocation policy, $H$. As  shown in Section \ref{sec:policies}, $H(\bm a_i|\bm X_i, M_i) = H(\bm a_i|\bm X_i, M_i;\pi)$ may depend on the cluster propensity score $\pi:=\pi(\bm a_i|\bm X_i, M_i)$. The value of $H(\bm a_i|\bm X_i, M_i)$ under the parametric submodel is $H(\bm a_i|\bm X_i, M_i;\pi_v)$ with $\pi_v:=\pi_v(\bm a_i|\bm X_i, M_i)=\mathbb{P}_v(\bm A_i=\bm a_i|\bm X_i, M_i)$. Following \cite{lee2024nonparametric}, we assume that a unique function $\Omega(\bm A_i, \bm X_i, M_i;\bm a_i)$ exists with $\E[\Omega(\bm A_i, \bm X_i, M_i;\bm a_i)|\bm X_i, M_i]=0$ and $\E[\Omega^2(\bm A_i, \bm X_i, M_i;\bm a_i)|\bm X_i, M_i]<\infty$ such that
\begin{equation}\label{eq:H_eif}
    \frac{d}{d v} H(\bm a_i'|\bm X_i, M_i;\pi_v) \Big|_{v=0} =  \E\left[ \Omega(\bm A_i, \bm X_i, M_i;\bm a_i') \mathcal S(\bm A_i, \bm X_i, M_i)|\bcx_i,M_i\right],
\end{equation}
where $\mathcal S(\bm A_i, \bm X_i, M_i)$ is the likelihood score of $\mathbb{P}_v(\bm A_i, \bm X_i, M_i)$. Condition \eqref{eq:H_eif} essentially assumes that, given $\{\bm X_i,M_i\}$, the EIF of $H$ exists under the nonparametric model of $\mathbb{P}(\bm A_i|\bm X_i,M_i)$. Therefore, we refer to $\Omega(\bm A_i, \bm X_i, M_i;\bm a_i)$  as the conditional EIF of $H(\bm a_i|\bm X_i, M_i)$ \citep{lee2024nonparametric}. It should be noted that, for the deterministic and uniform allocation policies, $H_a^{\texttt{DAP}}$ and $H_{\alpha}^{\texttt{UAP}}$ do not depend on data so that their corresponding $\Omega(\bm A_i, \bm X_i, M_i;\bm a_i)$ equals $0$. For the individual-level and cluster-level incremental propensity score policies, the conditional EIFs of $H_\delta^{\texttt{IPS}}$ and $H_{\delta}^{\texttt{CPS}}$ are $H_{\delta}^{\texttt{IPS}}(\bm a_i|\bm X_i, M_i) \displaystyle\sum_{j=1}^{M_i} \frac{(2a_{ij}-1)\delta (A_{ij}-\pi_{ij})}{{(\pi_{ij}^\delta)}^{a_{ij}}(1-\pi_{ij}^\delta)^{1-a_{ij}}[\delta \pi_{ij}+1-\pi_{ij}]^2}$ and $H_{\delta}^{\texttt{CPS}}(\bm a_i|\bm X_i, M_i) \Big\{\frac{\mathbb{I}(\bm A_i = \bm a_i)}{\pi(\bm a_i|\bm X_i, M_i)}  - \frac{H_{\delta}^{\texttt{CPS}}(\bm A_i|\bm X_i, M_i)}{\pi(\bm A_i|\bm X_i, M_i)}\Big\}$, respectively (detailed derivations are provided in Web Appendix A.2). 

Define $m_{ij}(\theta,\bm a_i,\bm X_i,M_i)
:= \mathbb{P}(Y_{ij}\le \theta| \bm A_i = \bm a_i,\bm X_i,M_i) = \mathbb{E}\!\left[\mathbb{I}(Y_{ij}\le \theta)| \bm A_i = \bm a_i,\bm X_i,M_i\right]$ as the conditional expectation of the threshold outcome $\mathbb{I}(Y_{ij}\le \theta)$ given $
\{\bm A_i=\bm a_i,\bm X_i,M_i\}$. We abbreviate it to $m(\theta)$, where the argument $\theta$ highlights that this conditional expectation depends on the threshold $\theta$. Proposition \ref{prop:EIF} suggests that $\{\pi,m(Q_{H}^{(t)})\} \equiv \{\pi(\bm a_i|\bm X_i,M_i),m_{ij}(Q_{H}^{(t)}, \\\bm a_i,\bm X_i,M_i)\}$ are two key nuisance parameters in the EIF of $Q_{H}^{(t)}\equiv Q_{H}^{(t)}(q)$. 


\begin{proposition}\label{prop:EIF}
(EIF) Suppose that Assumptions \ref{assum:super_population}–\ref{assum:unique} and condition \eqref{eq:H_eif} holds. Assume the conditional outcome density exists with $f_{Y_{ij}}(\theta|\bm A_i=\bm a_i,\bm X_i,M_i):=\frac{d}{d\theta}m_{ij}(\theta,\bm a_i,\bm X_i,M_i)>0$ for $\theta\in\Theta_t$. Then, for $t\in\{\star,0,1\}$ with any fixed $q\in \mathcal Q$, the EIF of $Q_H^{(t)} \equiv Q_H^{(t)}(q)$ is
\begin{align*}
\psi_H^{(t)}\left(\bm O_i;q\right) = \frac{1}{C_H^{(t)}(q)\cdot M_i}\sum_{j=1}^{M_i} \sum_{\bm a_i \in \mathcal A(M_i)}\left\{\psi_{ij}(\bm O_i;q,Q_H^{(t)},\pi) + \eta_{ij}\left(\bm O_i;Q_H^{(t)},\pi,m(Q_H^{(t)})\right)\right\},
\end{align*}
where $C_H^{(t)}(q) = \E\left[\frac{1}{M_i}\sum_{j=1}^{M_i} \sum_{\bm a_i \in \mathcal A(M_i)} w_{ij}^{(t)}(\bm a_i,\bm X_i,M_i) f_{Y_{ij}}(Q_H^{(t)}(q)|\bm A_i=\bm a_i,\bm X_i,M_i)\right]$, 
\begin{align*}
 \psi_{ij}(\bm O_i;q,Q_H^{(t)}(q),\pi) = &  \frac{ \mathbb{I}(\bm A_i = \bm a_i) w_{ij}^{(t)}(\bm a_i,\bm X_i,M_i)\{\mathbb{I}(Y_{ij}\leq Q_H^{(t)}(q) )-q\}}
     {\pi(\bm a_i | \bm X_i,M_i)}, \\
  \eta_{ij}\left(\bm O_i;Q_H^{(t)},\pi,m(Q_H^{(t)})\right) = &  \left\{\Omega_{ij}^{(t)}(\bm A_i, \bm X_i, M_i;\bm a_i) + w_{ij}^{(t)}(\bm a_i,\bm X_i,M_i)\frac{\pi(\bm a_i | \bm X_i,M_i)- \mathbb{I}(\bm A_i = \bm a_i) }{ \pi(\bm a_i | \bm X_i,M_i) }\right\}  \\
  & \quad \times\left\{m_{ij}\left(Q_H^{(t)}(q),\bm a_i,\bm X_i,M_i\right)-q\right\},
\end{align*}
with $\Omega_{ij}^{(\star)}(\bm A_i, \bm X_i, M_i;\bm a_i) = \Omega(\bm A_i, \bm X_i, M_i;\bm a_i)$ and $\Omega_{ij}^{(a)}(\bm A_i, \bm X_i, M_i;\bm a_i) = \mathbb{I}(a_{ij}=0)\sum_{a\in\{0,1\}}\\ \Omega(\bm A_i, \bm X_i, M_i;a,\bm a_{i(-j)})$ when $a\in\{0,1\}$. Consequently, the nonparametric efficient bound for estimating $Q_H^{(t)}(q)$ is $\E[\{\psi_H^{(t)}\left(\bm O_i,q\right)\}^2]$. 
\end{proposition}

In the EIF, both $w_{ij}^{(t)} \equiv w_{ij}^{(t)}(\bm a_i,\bm X_i, M_i)$ and $\Omega_{ij}^{(t)} \equiv \Omega_{ij}^{(t)}(\bm A_i, \bm X_i, M_i;\bm a_i)$ are either known explicitly (for the DAP and UAP policies) or depend on the cluster propensity score $\pi$ (for the IPS and CPS policies). To simplify notation, we do not explicitly display the dependence of $\psi_{ij}$ and $\eta_{ij}$ on $w_{ij}^{(t)}$ and $\Omega_{ij}^{(t)}$, and instead treat them as functions implicitly determined by $\pi$. Proposition \ref{prop:EIF} indicates that the EIF consists of three components: the IPW-type term $\psi_{ij}$, the augmentation term $\eta_{ij}$, and a normalizing constant $C_H^{(t)}(q)$ that depends only on the policy $H$ and the quantile level $q$. The component $\psi_{ij}$ corresponds to the IPW estimating function  in Theorem~\ref{thm: identification}. The second component $\eta_{ij}$ is an augmentation term involving the conditional mean of the threshold outcome $m_{ij}(Q_H^{(t)},\bm a_i, \bcx_i,M_i)$; the variance of the EIF characterizes the efficiency lower bound for estimating $Q_H^{(t)}(q)$. 


\subsection{Operational challenges for the EIF-based estimating equation}

Compared to IPW, an improved estimator of $Q_H^{(t)}(q)$ can be constructed by solving the following EIF-based estimating equation\footnote{For expositional simplicity, here we suppress cross-fitting procedures in the EIF-based estimating equations but will introduce it later.} in $\theta$:
$$
\frac{1}{n}\sum_{i=1}^n\frac{1}{C_H^{(t)}(q)\cdot M_i}\sum_{j=1}^{M_i}\sum_{\bm a_i \in \mathcal A(M_i)}
\Big\{\psi_{ij}(\bm O_i;q,\theta,\widehat \pi) + \eta_{ij}(\bm O_i;\widehat \pi,\widehat m(\theta))\Big\} = 0,
$$
where the target quantile estimand $Q_H^{(t)}(q)$ is replaced by the unknown parameter $\theta$, and the nuisance functions $\{\pi,m(\theta)\}$ are replaced by their estimated counterparts. Note that $C_H^{(t)}(q)$ need not be evaluated when solving the equation, since it does not affect the root. 

Operationalizing this direct plug-in estimator has two challenges. First, $m(\theta)$ depends on the parameter to be solved, meaning that evaluation of the estimating equation requires estimating $\{m(\theta): \theta\in\Theta_t\}$. In practice, this requires the estimation of the entire conditional distribution of $Y_{ij}$ given treatment, covariates, and cluster size, or, equivalently, infinitely many conditional expectations of $\mathbb{I}(Y_{ij}\leq \theta)$ indexed by thresholds $\theta$. This may be computationally burdensome and unstable when the covariate dimension is high. As one solution, we can discretize $\Theta_t$ based on a grid of points and then only use the estimate on the grid to approximate the entire $\{m(\theta): \theta\in\Theta_t\}$, but this can still be computationally intensive when the grid is dense and prone to discretization bias when it is coarse \citep{cheng2025inverting}. Second, although calculation of $C_H^{(t)}(q)$ is not required for point estimation, it is needed for variance estimation as it is a component for calculating the efficiency lower bound. Computing $C_H^{(t)}(q)$ requires estimating the conditional density of $Y_{ij}$ at $Q_H^{(t)}(q)$ given $(\bm A_i,\bm X_i,M_i)$, and multi-dimensional covariate-conditional density estimation is typically difficult and not always numerically stable in finite samples.

We introduce two practical modifications to address the aforementioned challenges. To tackle the first challenge, we leverage a three-way cross-fitting procedure \citep{kallus2024localized} to eliminate the need to estimate $m(\theta)$ for all $\theta \in \Theta_t$. Instead, we only estimate $m(\widehat Q_H^{(t),\texttt{ipw}}(q))$, where $\widehat Q_H^{(t),\texttt{ipw}}(q)$ is an initial IPW estimator of $Q_H^{(t)}(q)$ serving as a rough approximation to the target quantile estimand. Treating $\widehat Q_H^{(t),\texttt{ipw}}(q)$ as fixed, the estimation of $m(\widehat Q_H^{(t),\texttt{ipw}}(q)) = \mathbb{E}[\mathbb{I}(Y_{ij} \le \widehat Q_H^{(t),\texttt{ipw}}(q)) \mid \bm A_i=\bm a_i, \bm X_i, M_i]$ reduces to a binary regression problem, enabling the use of flexible machine learning methods designed for binary outcomes. To avoid overfitting, our cross-fitting procedure has three disjoint data folds: the first fold is used to estimate $\widehat Q_H^{(t),\texttt{ipw}}(q)$, the second to estimate $m(\widehat Q_H^{(t),\texttt{ipw}}(q))$, the pooled first and second folds to estimate $\pi$, and the third to evaluate the EIF-based estimating equation. We show that if the initial estimator $\widehat Q_H^{(t),\texttt{ipw}}(q)$ is sufficiently close to the truth at an $o_\Prob(n^{-1/4})$ rate, then the updated EIF-based estimator is $\sqrt{n}$-consistent, asymptotically normal, and nonparametrically efficient.

To address the second challenge, we employ a smoothed estimating equation approach that replaces the nonsmooth indicator function in the EIF, $\mathbb{I}(Y_{ij}\le \theta)$, with a smooth approximation $\Phi\left((\theta-Y_{ij})/h\right)$, where $\Phi(\cdot)$ is a continuously differentiable CDF (e.g., the standard normal CDF) and $h>0$ is a bandwidth parameter that controls the degree of smoothing (with $h\rightarrow0$ as $n\rightarrow\infty$). Such smoothed strategy have been used in prior work to overcome non-differentiability in quantile estimation problems \citep{he2023smoothed,cheng2024doubly}. By smoothing the EIF, the estimating equation becomes differentiable in $\theta$, enabling the use of derivative-based solvers to search for $\theta$ and facilitating analytical variance estimation via standard sandwich estimation. Notably, computation of standard errors requires no additional nuisance estimation beyond what is needed for point estimation, making the proposed procedure computationally more efficient.

\subsection{Proposed nonparametric efficient estimators}\label{sec:cross-fitting}

\begin{figure}[h]
\centering
\begin{tikzpicture}[
  >=Stealth,
  box/.style={draw, minimum width=1.8cm, minimum height=0.8cm, align=center},
  brace/.style={decorate, decoration={brace, amplitude=9pt}},
  sarrow/.style={->, decorate, decoration={snake, amplitude=0.6pt, segment length=5pt}}
]
\node[box] (D1) at (0,0)   {$\mathcal{O}_1$};
\node[box] (D2) at (2.5,0)   {$\mathcal{O}_2$};
\node[box] (D3) at (5,0)   {$\mathcal{O}_3$};
\node[box] (D4) at (7.5,0)   {$\mathcal{O}_4$};
\node[box] (D5) at (10,0)  {$\mathcal{O}_5$};

\node[below = 1.2 cm of D1] (psi) {calculate \eqref{eq:EIF_np}};
\draw[->] (0,-0.6) -- (psi.north);   

\draw[brace]  (5.9,-0.6) -- (1.6,-0.6) ;
\draw[sarrow] (3.75,-0.9) -- ++(0,-0.6)
      node[below] (IPW) {$\widehat{Q}_{H}^{(t),\texttt{ipw}}$};

\draw[brace]  (10.9,-0.6) -- (6.6,-0.6) ;
\draw[sarrow] (8.75,-0.9) -- ++(0,-0.6)
      node[below] (Mean) {$\widehat m(\widehat{Q}_{H}^{(t),\texttt{ipw}})$};

\draw[brace] (1.6,0.6) -- (10.9,0.6);
\draw[sarrow] (6.25,0.9) -- ++(0,0.6)
      node[above] (Pi) {$\{\widehat \pi,\widehat \Omega,\widehat H\}$};

\draw[->] (IPW.east) -- (Mean.west);

\draw[->] (Pi.west) -- ++(-7,0) -- ++(0,-3.85) -- (psi.west);
\draw[->] (Mean.south) -- ++(0,-0.3) -- ++(-8.75,0) -- (psi.south);
\end{tikzpicture}
\caption{Illustration of the three-way cross-fitting procedure with $L=5$ data folds.\label{fig:3-way}}
\end{figure}
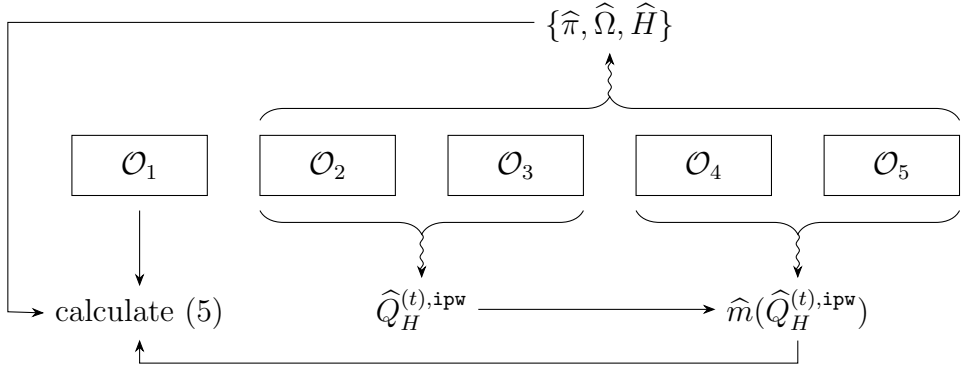

We propose the following three-way cross-fitting algorithm to estimate $Q_H^{(t)} \equiv Q_H^{(t)}(q)$, where an illustration of this procedure is given in Figure \ref{fig:3-way}.
\begin{compactitem}
\item[1.] Take a $L\geq 3$ fold random partition $\{\mathcal D_1,\dots,\mathcal D_L\}$ of data indices $\{1,\dots,n\}$ such that the size of each fold is approximately the same. For every $l\in\{1,\dots,L\}$, let $\mathcal O_l = \{\bm O_i: i\in \mathcal D_l\}$ be the observed data in $l$-th fold.  For each $l$, set $\mathcal I_{l,1} = \{\mathcal O_1,\dots,\mathcal O_{\lfloor 0.5L \rfloor+l'}\}/O_l$ and $\mathcal I_{l,2} = \{\mathcal O_{\lfloor 0.5L \rfloor+l'+1},\dots,\mathcal O_L\}/\mathcal O_l$, where $l' = \mathbb{I}(l\leq \lfloor 0.5L \rfloor)$. 
\item[2.] For every $l \in \{1,\dots,L\}$, do the following steps:
\begin{compactitem}
\item[a.] Only use $\mathcal I_{l,1}$ to estimate the initial IPW estimator $\widehat Q_{H,l}^{(t),\texttt{ipw}}$ based on procedures in Section \ref{sec:IPW}; 
\item[b.] Train a machine learning model based only on $\mathcal I_{l,2}$ to regress the binary outcome $\mathbb{I}(Y_{ij}\leq \widehat Q_{H,l}^{(t),\texttt{ipw}}(q))$ on $\{\bm A_i,\bm X_i, M_i\}$ to obtain $\widehat m_l(\widehat Q_{H,l}^{(t),\texttt{ipw}},\bm a_i, \bcx_i,M_i)$;
\item[c.] Fit a machine learning model (e.g., the hybrid copula model \eqref{eq:copula}) based on $\mathcal I_{l,1} \cup \mathcal I_{l,2}$ to obtain $\widehat \pi_l(\bm a_i|\bcx_i,M_i)$, and then calculate the associated $\widehat \Omega_l(\bm A_i,\bcx_i,M_i;\bm a_i)$ and $\widehat H_l(\bm a_i|\bcx_i,M_i)$ based on $\widehat \pi_l(\bm a_i|\bcx_i,M_i)$.  
\end{compactitem}
\item[3.] Finally, the proposed nonparametric efficient estimator, $\widehat Q_H^{(t),\texttt{np}}$, is obtained by solving the following estimating equation with repect to $\theta$:
\begin{equation}\label{eq:EIF_np}
\frac{1}{n}\sum_{l=1}^L\sum_{i \in D_l}\left[\frac{1}{M_i}\sum_{j=1}^{M_i}\sum_{\bm a_i \in \mathcal A(M_i)}
\Big\{\widetilde \psi_{ij}(\bm O_i;q,\theta,\widehat \pi_l) + \eta_{ij}\left(\bm O_i;\widehat \pi_l,\widehat m_l(\widehat Q_{H,l}^{(t),\texttt{ipw}})\right)\Big\}\right] = 0,
\end{equation}
where $\widetilde \psi_{ij}(\bm O_i;q,\theta,\pi)$ is a smoothed version of $\psi_{ij}(\bm O_i;q,\theta,\pi)$ that replaces the indicator function $\mathbb{I}(Y_{ij}\leq \theta)$ with $\Phi(\{\theta-Y_{ij}\}/h)$. 
\end{compactitem}

By construction, the parameter of interest in the estimating equation \eqref{eq:EIF_np}, $\theta$, is independent from the nuisance parameter  $\widehat m(\widehat Q_{H}^{(t),\texttt{ipw}})$, hence we obviate the need to estimate $m(\theta)$ for all possible $\theta$ in the direct EIF-based estimating equation. Our second modification introduces smoothness by replacing the indicator function $\mathbb{I}(Y_{ij}\le \theta)$ with its smoothed version $\Phi\{(\theta - Y_{ij})/h\}$. This makes the estimating equation differentiable in $\theta$ and enables inference using a sandwich variance estimator. Let $\gamma_{i}(\bm O_i,\theta)$ be the estimating score inside the square bracket in \eqref{eq:EIF_np}. The estimated variance of $\widehat Q_{H}^{(t),\texttt{np}}$ (scaled by $\sqrt{n}$) is given by
\begin{equation}\label{eq:variance_est}
\{\widehat \sigma_H^{(t)}(q)\}^2 :=  \widehat{\text{Var}}\{\sqrt{n}(\widehat Q_{H}^{(t),\texttt{np}})\} = \frac{1}{n}\sum_{l=1}^L\sum_{i \in D_l} \{\widehat C_H^{(t)}(q)\}^{-2} \left\{\gamma_{i}(\bm O_i,\widehat Q_{H}^{(t),\texttt{np}})\right\}^2,
\end{equation}
where 
$
\widehat C_H^{(t)}(q) = \frac{1}{n}\sum_{l=1}^L\sum_{i \in D_l} \frac{d}{d \theta} \gamma_{ij}(\bm O_i,\theta) \Big|_{\theta = \widehat Q_{H}^{(t),\texttt{np}}}
$
is the estimate of $C_H^{(t)}(q)$, which is defined as the empirical average of the derivative of the estimating score $\gamma_{i}(\bm O_i,\theta)$ with respect to $\theta$ at $\widehat Q_{H}^{(t),\texttt{np}}$. In our setting, $\widehat C_H^{(t)}(q)$ admits the closed-form expression
$$
\widehat C_H^{(t)}(q) = \frac{1}{nh}\sum_{l=1}^L\sum_{i \in D_l}\frac{1}{M_i}\sum_{j=1}^{M_i}\sum_{\bm a_i \in \mathcal A(M_i)} \frac{ \mathbb{I}(\bm A_i = \bm a_i) \widehat w_{ij,l}^{(t)}(\bm a_i,\bm X_i,M_i)\phi(\{\widehat Q_{H}^{(t),\texttt{np}}-Y_{ij}\}/h)}
     {\widehat \pi_l(\bm a_i | \bm X_i,M_i)},
$$ 
where $\phi(x) = \frac{d }{d x} \Phi(x)$. Importantly, computing $\widehat C_H^{(t)}(q)$ does not require estimating any additional nuisance components beyond those needed for point estimation. Finally, a Wald-type confidence interval for $\widehat Q_{H}^{(t),\texttt{np}}(q)$ can be constructed based on $\widehat \sigma_H^{(t)}(q)$ and normal quantiles.

\subsection{Large-sample properties}

This section studies the large-sample properties of $\widehat Q_{H}^{(t),\texttt{np}}$. 
We first introduce a few regularity conditions regarding the local smoothed function and the nuisance estimation.
\begin{assumption}\label{assum:smooth}
(a) The local CDF $\Phi(x)$ is continuous and its first-order derivative $\phi(x)$ is symmetric about zero with $\int x^2 \phi(x) dx <\infty$. (b) The bandwidth $h$ is chosen as $n\to \infty$, $nh\to \infty$ but $nh^4\to 0$. In other words, $h \asymp n^{-r}$ with $0.25<r<1$. 
\end{assumption}
We follow suggestions in \cite{heller2007smoothed} and \cite{cheng2024doubly} to set the bandwidth at $h=\widehat \sigma_Y \times (\sum_{i=1}^n M_i)^{-0.26}$, where $\widehat \sigma_Y$ is the empirical standard deviation of the observed outcome, $\sum_{i=1}^n M_i$ is the total number of individuals in the study, and the order $-0.26$ provides the quickest rate of convergence under the bandwidth constraint $nh^4 \to 0$ . 

\begin{assumption}\label{assum:convergence}
Assume that $\widehat Q_H^{(t),\texttt{np}}$ is an approximate solution to \eqref{eq:EIF_np} with an error of order $o_\Prob(n^{-1/2})$. Moreover, suppose that the following conditions hold for any $l\in\{1,\dots,L\}$.
\begin{compactitem}
\item[(a)] (Boundedness) For any $m_i\in \mathbb{M}$, $\bm x_i \in \mathcal X(m_i)$, and $\bm a_i, \bm a_i'\in \mathcal A(m_i)$, suppose that (i) $0<\widehat\pi_{l}(\bm a_i| \bm x_i, m_i)<1$; (ii) $|\widehat m_{ij,l}(\widehat Q_{H,l}^{(t),\texttt{ipw}},\bm a_i, \bm x_i, m_i)|<\infty$. (iii) $|\widehat H_{l}(\bm a_i|\bm x_i, m_i)|<\infty$. (iv) $|\Omega(\bm a_i, \bm x_i, m_i;\bm a_i')|<\infty$ and $|\widehat \Omega_l(\bm a_i, \bm x_i, m_i;\bm a_i')|<\infty$.
\item[(b)] (Smoothness of conditional outcome distribution) Assume 
$
\left\lvert \frac{d}{d\theta} m_{ij}(\theta,\bm a_i, \bm x_i, m_i)\right\rvert < \infty, \quad \left\lvert \frac{d^2}{d\theta^2} m_{ij}(\theta,\bm a_i, \bm x_i, m_i)\right\rvert < \infty
$
uniformly in $\theta \in \mathcal Y$, $m_i\in \mathbb{M}$, $\bm x_i \in \mathcal X(m_i)$, and $\bm a_i \in \mathcal A(m_i)$.
\item[(c)] (Convergence rates). For the initial IPW estimator,  
$
\sup_{q\in\mathcal Q}|\widehat Q_{H,l}^{(t),\texttt{ipw}} - Q_H^{(t)}| =  O_\Prob(r_\theta)
$ for some $r_{\theta}>0$. For nuisance functions, there exist $r_\pi, r_H, r_m, r_\Omega>0$ such that
\begin{align}
& \left\|\sum_{\bm a_i \in \mathcal A(M_i)}|(\widehat\pi_{l} - \pi)(\bm a_i, \bcx_i, M_i)|\right\|_{2} = O_\Prob(r_\pi), \nonumber \\
& \sup_{\theta \in \Theta_t }\left\|\sum_{\bm a_i \in \mathcal A(M_i)}|(\widehat m_{ij,l} - m_{ij})(\theta,\bm a_i, \bcx_i, M_i;\theta)|\right\|_{2}  = O_\Prob(r_m) \nonumber \\
& \left\|\sum_{\bm a_i \in \mathcal A(M_i)}|(\widehat \Omega_{l}- \Omega)(\bca_i, \bcx_i, M_i;\bm a_i)|\right\|_{2}  = O_\Prob(r_\Omega) \nonumber \\
& \left\|\sum_{\bm a_i \in \mathcal A(M_i)}|(\widehat H_{l} \!-\! H)(\bm a_i| \bcx_i, M_i)+\E[\widehat \Omega_{-l}(\bca_i, \bcx_i, M_i;\bm a_i)|\bcx_i, M_i]|\right\|_{2} = O_\Prob(r_H^2).\label{eq:convergence_order}
\end{align}
\end{compactitem}
\end{assumption}

Assumption \ref{assum:convergence}(a) bounds the nuisance functions and their estimators, while Assumption \ref{assum:convergence}(b) requires the conditional distribution of observed outcome $m_{ij}(\theta,\bm a_i, \bm x_i, m_i)$ to be sufficiently smooth. Assumption \ref{assum:convergence}(c) imposes convergence rate requirements on the initial IPW estimator $\widehat Q_{H}^{(t),\texttt{ipw}}$ and the nuisance components $\{\widehat \pi,\widehat m(\widehat Q_{H}^{(t),\texttt{ipw}}),\widehat H,\widehat \Omega\}$. Here, $r_\theta$ and $r_\pi$ are the convergence rates of the IPW estimator $\widehat Q_{H}^{(t),\texttt{ipw}}$ and the cluster propensity score estimator; as shown in Proposition \ref{thm:ipw}, for all polices in Section \ref{sec:policies}, $r_\theta = O(r_\pi+n^{-1/2})$. 
Under the $L_2(\mathbb{P})$ norm, $r_m$ denotes the convergence rate of machine learning model of regressing of $\mathbb{I}(Y_{ij}\leq \theta)$ on $\{\bm A_i, \bm X_i, M_i\}$; $r_\Omega$ denotes the convergence rate of the estimated local EIF component of the policy-specific treatment allocation probability $\widehat\Omega$. Following \cite{lee2025efficient}, the condition in \eqref{eq:convergence_order} characterizes the convergence rate of the second-order remainder term in the von Mises expansion \citep{kennedy2024semiparametric} of $H(\bm a_i \mid \bm X_i, M_i)$. This second-order term typically converges at a rate proportional to the square of the first-order error $(\widehat H_l - H)(\bm a_i \mid \bm X_i, M_i)$, and we therefore write its rate as $r_H^2$. Under these regularity conditions, we now establish the asymptotic behavior of $\widehat Q_{H}^{(t),\texttt{np}}$:

\begin{theorem}\label{thm:np}
Suppose that Assumptions \ref{assum:super_population}--\ref{assum:convergence} hold. 
\begin{compactitem}
\item[a.] ($\sqrt{n}$-consistency and asymptotic normality) 
If (i) $r_\theta = r_\pi = r_m = r_{\Omega} = o(1)$, (ii) $r_H=o(n^{-1/4})$, and (iii) $(r_m+r_\theta)(r_\pi+r_H) = o(n^{-1/2})$ as $n\to \infty$, then we have that 
$$
\sqrt{n} \left\{\widehat Q_{H}^{(t),\texttt{np}}(q) - Q_{H}^{(t)}(q)\right\} \overset{d}{\to} N\left(0,\{\sigma_H^{(t)}(q)\}^2\right),
$$
where $\{\sigma_H^{(t)}(q) \}^2= \E[\{\psi_H^{(t)}\left(\bm O_i\right)\}^2]$ is the nonparametric efficiency bound of $Q_{H}^{(t)}(q)$.
\item[b.] (Consistent standard error estimation) 
If $r_\theta = r_\pi = r_m = r_{\Omega} = r_H = o(1)$ as $n\to \infty$, then $\widehat \sigma_H^{(t)}(q)$ defined in \eqref{eq:variance_est} is consistent such that $\widehat \sigma_H^{(t)}(q) = \sigma_H^{(t)}(q) + o_\Prob(1)$.  
\end{compactitem}
\end{theorem}

As stated in Theorem \ref{thm:np}(a), the rate conditions of the nuisance estimates are slower than the parametric convergence rate. Specifically, it is sufficient to show if both (i) $r_\theta = r_\pi = r_m = r_{H} = o(n^{-1/4})$ and (i) $r_\Omega=o(1)$ as $n\to \infty$, then $\widehat Q_{H}^{(t),\texttt{np}}$ is $\sqrt{n}$-consistent, asymptotically normal, with its asymptotic variance achieving the nonparametric efficiency bound. Theorem \ref{thm:np}(c) suggests that, if all nuisance functions are consistently estimated with $r_\theta = r_\pi = r_m = r_{H} = r_\Omega = o(1)$ as $n\to\infty$, the standard error estimation $\widehat \sigma_H^{(t)}(q)$ is consistent to the true value $\sigma_H^{(t)}(q)$, thus providing a  basis for asymptotic inference. 


In Proposition \ref{prop:np}, we provide sufficient conditions for each example policy outlined in Section~\ref{sec:policies} under which the assumptions in Theorem~2(a)--(b) are satisfied. For each policy, Proposition~\ref{thm:ipw} implies that $r_\theta = O(r_\pi+n^{-1/2})$. 
Moreover, for all example policies, the estimators $\Omega$ and $H$ are either known explicitly (for the DAP and UAP policies) or constructed solely from the estimated cluster propensity score $\pi$ (for the IPS and CPS policies). Consequently, the convergence rates $\{r_\Omega, r_H\}$ are governed by the convergence rate of $\widehat \pi$, namely $r_\pi$. In particular, as shown in Appendix~A.2, we establish that $r_\Omega = O(r_\pi)$ and $r_H = O(r_\pi)$ for the IPS and CPS policies, whereas $r_\Omega = r_H = 0$ for the DAP and UAP policies. These observations imply that the asymptotic behavior of $\widehat Q_H^{(t),\texttt{np}}$ under each example policy is fully characterized by the convergence rates $\{r_\pi, r_m\}$, corresponding to the estimation of cluster propensity score and conditional outcome regression.

\begin{proposition}\label{prop:np}
For all policies in Section \ref{sec:policies}, we have the following results:
\begin{compactitem}
\item[a.] if $r_\pi=o(n^{-1/4})$ and $r_m=o(n^{-1/4})$, then conditions in Theorem \ref{thm:np}(a) are satisfied. 
\item[b.] if $r_\pi=r_m=o(1)$, then conditions in Theorem \ref{thm:np}(b) are satisfied. 
\end{compactitem}
\end{proposition}


As is shown in Proposition \ref{prop:np}, for all example policies, if $\widehat\pi$ and $\widehat m(\theta)$ converge to their population counterparts at rate $o_\Prob(n^{-1/4})$ under the $L_2(\mathbb{P})$ norm, then the resulting estimator $\widehat Q_H^{(t),\texttt{np}}$ is $\sqrt{n}$-consistent, asymptotically normal, and achieves the semiparametric efficiency bound. Moreover, the proposed standard error estimator $\widehat \sigma_H^{(t)}(q)$ is consistent.

\subsection{Weak convergences over quantiles}\label{sec:convergence_quantiles}

When we are interested in studying how network treatment effect varies over quantile level $q\in\mathcal Q =[\epsilon, 1-\epsilon]$, Theorem \ref{thm:np}(a) can be generalized to settings when the estimand depends on the quantile level $q$. The following theorem shows that the proposed nonparametric efficient estimator, $\sqrt{n}\left\{\widehat Q_H^{(t),\texttt{np}}(q)-Q_H^{(t)}(q)\right\}\Big/\widehat \sigma_H^{(t)}(q)$, weakly converges to a zero-mean Gaussian process, as in the standard quantile process analysis. Let $l^{\infty}(\mathcal Q)$ denote the function space with the supermum norm $\|f\|_{\mathcal Q} = \sup_{q\in \mathcal Q} | f(q)|$. 

\begin{theorem}\label{thm:np_weak_convergence}
Suppose that Assumptions \ref{assum:super_population}--\ref{assum:convergence} hold, and all rate of convergence conditions in Theorem \ref{thm:np}(a) hold. Then, we have that
$
\sqrt{n}\left\{\widehat Q_H^{(t),\texttt{np}}(\cdot)-Q_H^{(t)}(\cdot)\right\}\Big/\widehat \sigma_H^{(t)}(\cdot) \leadsto \mathbb{G}(\cdot)
$
in $l^{\infty}(\mathcal Q)$ as $n\to \infty$, where $\mathbb{G}(\cdot)$ is a mean zero Gaussian process with covariance $\E[\mathbb{G}(q_1)\mathbb{G}(q_2)] = \text{Cov}\left\{\psi_H^{(t)}\left(\bm O_i,q_1\right)/\sigma_H^{(t)}(q_1),\psi_H^{(t)}\left(\bm O_i,q_2\right)/\sigma_H^{(t)}(q_2)\right\}$, and $\psi_H^{(t)}\left(\bm O_i,q\right)$ is the EIF of $Q_H^{(t)}(q)$. 
\end{theorem}

Based on Theorem \ref{thm:np_weak_convergence}, one can construct uniform confidence bands for the counterfactual quantile curve $\{Q_H^{(t)}(q): q\in \mathcal Q\}$. We aim to construct a $(1-\alpha)$ uniform confidence band with the form $\widehat Q_H^{(t),\texttt{np}}(q) \pm c_\alpha \widehat \sigma_H^{(t)}(q)/\sqrt{n}$, where $c_\alpha$ is the critical value that satisfies
\begin{equation}\label{eq:supremum}
\Prob \left[\sup_{q\in\mathcal Q} \left| \frac{\widehat Q_H^{(t),\texttt{np}}(q) -  Q_H^{(t)}(q)}{\widehat \sigma_H^{(t)}(q)/\sqrt{n}} \right| \leq c_\alpha \right]  = 1-\alpha + o(1). 
\end{equation}
For implementation, we pursue the multiplier bootstrap approach \citep{belloni2018uniformly,kennedy2019nonparametric} to construct the uniform confidence band, because it is computationally convenient without the need to refit the nuisance estimators. We shall approximate the supremum in \eqref{eq:supremum} based on the following supremum of the  multiplier process
\begin{equation}\label{eq:sample_supremum}
\sup_{q\in\mathcal Q} \left|\frac{1}{\sqrt{n}}\sum_{l=1}^L\sum_{i \in D_l} \xi_i \widehat \psi_H^{(t)}\left(\bm O_i,q\right)/\widehat \sigma_H^{(t)}(q)\right|
\end{equation}
where $\widehat \psi_H^{(t)}\left(\bm O_i,q\right)$ is the estimated EIF of $\widehat Q_H^{(t),\texttt{np}}(q)$. Here, $\{\xi_1,\dots,\xi_n\}$ are $n$ independent draws from the Rademacher multiplier (i.e., $\mathbb{P}(\xi=1) = \mathbb{P}(\xi=-1) = 0.5$). The critical value $\widehat c_\alpha$ is chosen as the $(1-\alpha)$ sample quantile of the supremum of the multiplier process \eqref{eq:sample_supremum}. Finally, the $(1-\alpha)$ uniform confidence band is given by $\widehat Q_H^{(t),\texttt{np}}(q) \pm \widehat c_\alpha \widehat \sigma_H^{(t)}(q)/\sqrt{n}$.  Proposition \ref{prop:bootstrap} establishes a theoretical justification for the multiplier bootstrap approach. 
\begin{proposition}\label{prop:bootstrap}
Under the same conditions in Theorem \ref{thm:np_weak_convergence}, we have that
$$
\underset{n\to \infty}{\text{lim}}\Prob \left[\sup_{q\in\mathcal Q} \left| \frac{\widehat Q_H^{(t),\texttt{np}}(q) -  Q_H^{(t)}(q)}{\widehat \sigma_H^{(t)}(q)/\sqrt{n}} \right| \leq \widehat c_\alpha \right]  = 1-\alpha. 
$$
\end{proposition}

\subsection{Weak convergences over policies}

In previous analyses, our results in Theorem \ref{thm:np} is based on a fixed policy $H$. Now, consider we have a collection of polices $H_{\beta}$ indexed by parameter $\beta \in \mathcal B$. Then, we can generalize our results from a point-wise analysis with a fixed $\beta$ to uniform analysis on $\{Q_{H,\beta}^{(t)}(q):\beta\in \mathcal B\}$. Let $l^{\infty}(\mathcal B)$ denote the function space with the supermum norm $\|f\|_{\mathcal B} = \sup_{\beta\in \mathcal B} | f(\beta)|$. In what follows, we develop the weak convergence results for the cluster-level incremental propensity score policy $H_{\delta}^{\texttt{CPS}}$ with $\delta \in \Delta :=[\underline \delta,\overline \delta]$ (with $0<\underline \delta<\overline \delta<\infty$).



\begin{theorem}\label{thm:weak_convergence_CPS}
Suppose that Assumptions \ref{assum:super_population}--\ref{assum:convergence} hold with $r_\pi=o(n^{-1/4})$ and $r_m=o(n^{-1/4})$. Consider the collection of $H_{\delta}^{\texttt{ICP}}$ with $\delta \in \Delta :=[\underline \delta,\overline \delta]$ with $0<\underline \delta<\overline \delta<\infty$. For a fixed $q\in\mathcal Q$, 
$
\sqrt{n}\left\{\widehat Q_{\texttt{CPS},\delta}^{(t),\texttt{np}}(q)-Q_{\texttt{CPS},\delta}^{(t)}(q)\right\}\Big/\widehat \sigma_{\texttt{CPS},\delta}^{(t)}(q) \leadsto \mathbb{G}(\cdot)
$
in $l^{\infty}(\Delta)$ as $n\to \infty$, where $\mathbb{G}(\cdot)$ is a mean zero Gaussian process with covariance $\E[\mathbb{G}(\delta_1)\mathbb{G}(\delta_2)] = \text{Cov}\Big\{\psi_{\texttt{CPS},\delta_1}^{(t)}\left(\bm O_i,q\right)/\sigma_{\texttt{CPS},\delta_1}^{(t)}(q),\\ \psi_{\texttt{CPS},\delta_2}^{(t)}\left(\bm O_i,q \right)/\sigma_{\texttt{CPS},\delta_2}^{(t)}(q)\Big\}$, and $\psi_{\texttt{CPS},\delta}^{(t)}\left(\bm O_i,q\right)$ is the EIF of $Q_{\texttt{CPS},\delta}^{(t)}(q)$. 
\end{theorem}

Under the same rate conditions required in Proposition \ref{prop:np}(a), Theorem \ref{thm:weak_convergence_CPS} show that the nonparametric efficient estimator $\{\sqrt{n}(\widehat Q_{\texttt{CPS},\delta}^{(t)}(q)-Q_{\texttt{CPS},\delta}^{(t)}(q))/\widehat \sigma_{\texttt{CPS},\delta}^{(t)}(q):\delta \in \Delta\}$ weakly converges to a mean zero Gaussian process. Based on the multiplier bootstrap approach shown in Section \ref{sec:convergence_quantiles}, one can similarly construct a $(1-\alpha)$ uniform confidence band over the policies index range. The multiplier bootstrap procedure is omitted for brevity.

Similar to Theorem \ref{thm:weak_convergence_CPS}, Web Appendix A.3 develops the weak convergence results for the UAP $H_{\alpha}^{\texttt{UAP}}$ with $\alpha\in [\underline \alpha,\overline \alpha]$ ($0<\underline \alpha<\overline \alpha<1$) and the IPS $H_{\delta}^{\texttt{IPS}}$ with $\delta \in [\underline \delta,\overline \delta]$ ($0<\underline \delta<\overline \delta<\infty$). There is no weak convergence result for the deterministic allocation policy since its parameter $a\in\{0,1\}$ is discrete. 

\subsection{Estimation of network quantile causal effects}

After obtaining $\widehat Q_H^{(t),\texttt{np}}(q)$, the network quantile causal effects can be directly calculated by
\begin{align*}
\widehat{\text{OQE}}^{\texttt{np}}(q; H, H') = \widehat Q_H^{(\star),\texttt{np}}(q) - \widehat Q_{H'}^{(\star),\texttt{np}}(q), \quad & \widehat{\text{DQE}}^{\texttt{np}}(q; H) =\widehat Q_H^{(1),\texttt{np}}(q) - \widehat Q_H^{(0),\texttt{np}}(q), \\
\widehat{\text{SQE}}_a^{\texttt{np}}(q; H,H') = \widehat Q_H^{(a),\texttt{np}}(q) - \widehat Q_{H'}^{(a),\texttt{np}}(q), \quad & \widehat{\text{TQE}}^{\texttt{np}}(q; H, H') = \widehat Q_H^{(1),\texttt{np}}(q) - \widehat Q_{H'}^{(0),\texttt{np}}(q). 
\end{align*}
Because these estimators are linear combinations of the estimated quantile curves, their asymptotic properties follow directly from Theorems \ref{thm:np}–\ref{thm:weak_convergence_CPS}. In particular, consistency, asymptotic normality, and weak convergence for the network quantile causal effects are immediate consequences of the corresponding results for $Q_H^{(t)}(q)$. Inference for the network quantile causal effects can therefore be conducted using the same EIF representation and multiplier bootstrap procedure developed for the policy-specific quantile estimands.

\section{Simulation studies}\label{sec:sim}

We conduct simulation studies to evaluate  performance of the proposed estimators. We generate 1{,}000 clustered observational studies with each $n=500$ clusters. For each simulated study, we generate the full data $\mathcal W_i = \{M_i,\bm X_i = \{\bm X_{i,1},\bm X_{i,2}, \bm X_{i,3}\}, \bm A_i, \bm Y_i(\bullet )\}$ as follows. First, we draw the cluster size $M_i \in \text{Unif}\{3,4,5,6\}$. Then, for each $j \in \{1,\dots, M_i\}$, we generate three mutually independent covariates $X_{ij,1} \sim N(0,1)$, $X_{ij,2} \sim N(0,1)$, and $X_{ij,3} \sim \text{Bernoulli}(0.5)$.  To induce within-cluster correlation while preserving marginal distributions, we generate $\bm A_i$ based a hybrid copula model in \eqref{eq:copula}. Specifically,  $\bm A_i=[A_{i1},\dots,A_{iM_i}]^T$ are sampled from the following Gaussian copula model:
$$
\mathbb{P}(A_{i1}\le a_{i1},\dots,A_{iM_i}\le a_{iM_i} | \bm X_i, M_i) =
\Phi_{M_i}\!\left(
\Phi^{-1}(\Pi_{i1}(a_{i1})),\dots,
\Phi^{-1}(\Pi_{iN_i}(a_{iM_i}))
\right),
$$
where $\Phi^{-1}$ denotes the inverse standard normal CDF, and $\Phi_n$ is the CDF of an 
$n$-variate standard normal distribution with an exchangeable correlation structure 
(pairwise correlation $0.1$). The marginal CDF $\Pi_{ij}(a_{ij})=\mathbb{P}(A_{ij}\le a_{ij}|\bm X_i, M_i)$ follows $\mathbb{P}(A_{ij}=1|\bm X_i,M_i) =\text{expit}\!\left(-0.5 + 0.3X_{ij,1} + 0.3X_{ij,2} + 0.3X_{ij,3}\right)$. Similarly, $\bm Y_i(\bm a_i)=[Y_{i1}(\bm a_i),\dots,Y_{iM_i}(\bm a_i)]^T$ are sampled from
$$
\mathbb{P}(Y_{i1}(\bm a_i)\le y_{i1},\dots,Y_{iM_i}(\bm a_i)\le y_{iM_i} \mid \bm A_i,\bm X_i,M_i)
= \Phi_{M_i}\!\left(
\Phi^{-1}(\mathcal P_{i1}(y_{i1})),\dots,
\Phi^{-1}(\mathcal P_{iM_i}(y_{iM_i}))
\right),
$$
where marginally $\mathcal P_{ij}(y_{ij})=\mathbb{P}(Y_{ij}(\bm a)\le y_{ij}| \bm A_i,\bm X_i,M_i)$ with $Y_{ij}(\bm a_i)|\{\bm A_i,\bm X_i,M_i\}\sim N\Big(3+1.5a_{ij}+3\frac{1}{M_i-1}\sum_{l\neq j}a_{il}+5X_{ij,1} + 5X_{ij,2} + 2X_{ij,3},1+a_{ij}\Big)$. Finally, the observed data $\bm O_i$ is extracted from the full data $\mathcal W_i$ by  $\bm O_i = \{M_i,\bm X_i, \bm A_i, \bm Y_i = \bm Y_i(\bm A_i)\}$.  In this simulation, we target the counterfactual quantile estimands $Q_H^{(\star)}(q)$, $Q_H^{(1)}(q)$, and $Q_H^{(0)}(q)$ under the cluster-level incremental propensity score policy (i.e., $H_{\delta}^{\texttt{CPS}}$) with $\delta$ chosen from $\{0.5,1,2\}$. The corresponding true values are computed based on Definition \ref{def:quantile} from a simulated super-population with known potential outcomes. We shall evaluate their bias, Monte Carlo standard deviation, and 95\% confidence interval coverage over 1{,}000 simulation replicates. 

We compare the performance of IPW and nonparametric efficient estimators for the three counterfactual quantile estimands, which require modeling treatment mechanism $\pi(\bm a_i|\bm X_i,M_i)$ and outcome regression $m(\widehat Q_H^{(t),\texttt{ipw}},\bm a_i,\bm X_i,M_i)$. Here, $\pi(\bm a_i|\bm X_i,M_i)$ is estimated based on the hybrid copula model \eqref{eq:copula}, where $\mathcal C_{\bm \gamma}$ is chosen as the Gaussian copula with exchangeable correlation matrix. The individual propensity score $\pi_{ij}(a_{ij}|\bm X_i, M_i)$ is estimated by regressing $A_{ij}$ on the feature matrix $\{ X_{ij,1}, X_{ij,2}, X_{ij,3},M_i\}$, based on the \texttt{SuperLearner} package in \texttt{R} software incorporating the generalized linear model, random forest, extreme gradient boosting, and generalized additive model libraries \citep{phillips2023practical}. For $m(\widehat Q_H^{(t),\texttt{ipw}},\bm a_i,\bm X_i,M_i)$, we regress the binary outcome $\mathbb{I}(Y_{ij} \leq \widehat Q_H^{(t),\texttt{ipw}})$ on the feature matrix $\{A_{ij}, \frac{1}{M_i-1}\sum_{l\neq j} A_{il}, X_{ij,1}, X_{ij,2}, X_{ij,3},M_i\}$ based on Super Learner with the same machine learner libraries. Although machine learners are more flexible than parametric models for estimating the nuisance functions, its performance still depends on the feature representation and may introduce finite-sample bias due to misspecification of feature matrices. Therefore, to evaluate the robustness of the proposed estimators under model misspecification, we consider two simulation scenarios. In Scenario (A), we use the aforementioned feature matrices for Super Learner models;  In Scenario (B), we construct two transformed covariates $U_{ij,1}=\exp(-0.5X_{ij,1})$, $U_{ij,2}=X_{ij,1}/(1+0.5X_{ij,2})$ and replace $\{X_{ij,1},X_{ij,2},X_{ij,3}\}$ in feature matrices with $\{U_{ij,1},U_{ij,2},X_{ij,3}\}$.

We begin by examining the quantile estimands at $q=0.5$. Figure \ref{fig:sim_50_A} reports the bias, Monte Carlo standard deviation, and coverage of $\widehat Q_{\texttt{CPS}}^{(t),\texttt{ipw}}$ and $\widehat Q_{\texttt{CPS}}^{(t),\texttt{np}}$ under Scenario (A), where both nuisance models are correctly specified. Both estimators present negligible bias and coverage close to the nominal level, though the proposed $\widehat Q_{\texttt{CPS}}^{(t),\texttt{np}}$ is notably more efficient, with RMSE ratios ranging from 0.77 to 1.00 across values of $\delta$. We next assess performance under model misspecification. As demonstrated in Figure \ref{fig:sim_50_B}, both $\widehat Q_{\texttt{CPS}}^{(t),\texttt{ipw}}$ and $\widehat Q_{\texttt{CPS}}^{(t),\texttt{np}}$ exhibit bias and reduced coverage, but the nonparametric efficient estimator shows a marked improvement, with much smaller  bias than the IPW estimator. In addition, $\widehat Q_{\texttt{CPS}}^{(t),\texttt{np}}$ achieves more stable coverage, ranging between 87\% and 95\%, whereas that of $\widehat Q_{\texttt{CPS}}^{(t),\texttt{ipw}}$ at times falls to around 50\%. The RMSE ratios range from 0.49 to 1.00 across values of $\delta$, favoring $\widehat Q_{\texttt{CPS}}^{(t),\texttt{np}}$. We also evaluate performance at lower ($q=0.2$) and higher ($q=0.8$) quantiles, with results provided in Supplementary Material Tables 1--4, which show qualitatively similar patterns to those observed at $q=0.5$.

\begin{table}[ht]
\centering
\caption{Simulation results for $Q_{\texttt{CPS}}^{(\star)}(0.5)$, $Q_{\texttt{CPS}}^{(1)}(0.5)$, and $Q_{\texttt{CPS}}^{(0)}(0.5)$ in Scenario (A).\label{fig:sim_50_A}}
\renewcommand{\arraystretch}{1.2}
\vspace{-0.3cm}
\scalebox{0.86}{ \begin{threeparttable}
\begin{tabular}{cccccccccc}
  \hline
  \multirow{2}{*}{$\delta$} & \multirow{2}{*}{Estimand} & \multirow{2}{*}{Truth} & \multicolumn{3}{c}{IPW} & \multicolumn{3}{c}{Nonparametric } & RMSE\\
\cmidrule(lr){4-6} \cmidrule(lr){7-9}
   &  &  & Bias & MCSD & Coverage & Bias & MCSD & Coverage & Ratio \\ 
  \hline
 0.5 & $Q_{\texttt{CPS}}^{(\star)}$ & $5.072$ & $-0.002$ & $0.251$ & $95.0$ & $-0.001$ & $0.222$ & $95.3$ & $0.884$ \\
 & $Q_{\texttt{CPS}}^{(1)}$ & $6.253$ & $-0.030$ & $0.354$ & $97.9$ & $-0.014$ & $0.280$ & $94.5$ & $0.789$ \\
 & $Q_{\texttt{CPS}}^{(0)}$ & $4.756$ & $0.008$ & $0.270$ & $96.5$ & $0.001$ & $0.226$ & $94.0$ & $0.836$ \\
1 & $Q_{\texttt{CPS}}^{(\star)}$ & $5.831$ & $-0.006$ & $0.224$ & $94.7$ & $-0.005$ & $0.225$ & $94.6$ & $1.004$ \\
 & $Q_{\texttt{CPS}}^{(1)}$ & $6.761$ & $-0.031$ & $0.272$ & $98.2$ & $-0.013$ & $0.264$ & $93.3$ & $0.965$ \\
 & $Q_{\texttt{CPS}}^{(0)}$ & $5.264$ & $0.011$ & $0.232$ & $97.9$ & $-0.000$ & $0.219$ & $94.5$ & $0.945$ \\
2 & $Q_{\texttt{CPS}}^{(\star)}$ & $6.724$ & $-0.010$ & $0.286$ & $93.6$ & $-0.010$ & $0.262$ & $93.3$ & $0.916$ \\
 & $Q_{\texttt{CPS}}^{(1)}$ & $7.343$ & $-0.030$ & $0.339$ & $97.2$ & $-0.012$ & $0.296$ & $93.6$ & $0.872$ \\
 & $Q_{\texttt{CPS}}^{(0)}$ & $5.846$ & $0.012$ & $0.313$ & $97.4$ & $0.000$ & $0.240$ & $94.4$ & $0.767$ \\
   \hline
\end{tabular}
\begin{spacing}{1}
\begin{tablenotes}
      \item[*] Truth: true value of the estimand. MCSD: Monte Carlo standard deviation. Coverage: 95\% confidence interval coverage. RMSE ratio: ratio of the root mean squared error (RMSE) between $\widehat Q_{\texttt{CPS}}^{(t),\texttt{np}}$ and $\widehat Q_{\texttt{CPS}}^{(t),\texttt{ipw}}$ (values $<1$ indicate higher efficiency of $\widehat Q_{\texttt{CPS}}^{(t),\texttt{np}}$). Since there is no established asymptotic distribution theory for the IPW estimator, we construct its confidence interval using an \textit{ad hoc} method (see Appendix~A.4 for details).
    \end{tablenotes}
\end{spacing}
\end{threeparttable}}
\end{table}

\begin{table}[ht]
\centering
\caption{Simulation results for $Q_{\texttt{CPS}}^{(\star)}(0.5)$, $Q_{\texttt{CPS}}^{(1)}(0.5)$, and $Q_{\texttt{CPS}}^{(0)}(0.5)$ in Scenario (B). \label{fig:sim_50_B}}
\renewcommand{\arraystretch}{1.2}
\vspace{-0.3cm}
\scalebox{0.86}{ \begin{threeparttable}
\begin{tabular}{cccccccccc}
  \hline
  \multirow{2}{*}{$\delta$} & \multirow{2}{*}{Estimand} & \multirow{2}{*}{Truth} & \multicolumn{3}{c}{IPW} & \multicolumn{3}{c}{Nonparametric } & RMSE\\
\cmidrule(lr){4-6} \cmidrule(lr){7-9}
   &  &  & Bias & MCSD & Coverage & Bias & MCSD & Coverage & Ratio \\ 
  \hline
 0.5 & $Q_{\texttt{CPS}}^{(\star)}$ & $5.072$ & $-0.223$ & $0.250$ & $87.8$ & $-0.077$ & $0.224$ & $93.7$ & $0.707$ \\
 & $Q_{\texttt{CPS}}^{(1)}$ & $6.253$ & $0.432$ & $1.420$ & $65.7$ & $0.084$ & $0.807$ & $90.0$ & $0.547$ \\
 & $Q_{\texttt{CPS}}^{(0)}$ & $4.756$ & $-0.462$ & $1.357$ & $55.9$ & $-0.159$ & $0.768$ & $89.2$ & $0.547$ \\
1 & $Q_{\texttt{CPS}}^{(\star)}$ & $5.831$ & $-0.032$ & $0.218$ & $94.1$ & $-0.029$ & $0.220$ & $94.2$ & $1.004$ \\
 & $Q_{\texttt{CPS}}^{(1)}$ & $6.761$ & $0.401$ & $1.414$ & $58.7$ & $0.081$ & $0.740$ & $90.8$ & $0.507$ \\
 & $Q_{\texttt{CPS}}^{(0)}$ & $5.264$ & $-0.503$ & $1.343$ & $49.7$ & $-0.157$ & $0.796$ & $87.6$ & $0.566$ \\
2 & $Q_{\texttt{CPS}}^{(\star)}$ & $6.724$ & $0.185$ & $0.279$ & $91.3$ & $0.021$ & $0.250$ & $95.0$ & $0.751$ \\
 & $Q_{\texttt{CPS}}^{(1)}$ & $7.343$ & $0.388$ & $1.424$ & $68.0$ & $0.085$ & $0.729$ & $91.3$ & $0.497$ \\
 & $Q_{\texttt{CPS}}^{(0)}$ & $5.846$ & $-0.542$ & $1.387$ & $64.2$ & $-0.155$ & $0.815$ & $89.5$ & $0.557$ \\
   \hline
\end{tabular}
\begin{spacing}{1}
\begin{tablenotes}
      \item[*] Truth: true value of the estimand. MCSD: Monte Carlo standard deviation. Coverage: 95\% confidence interval coverage. RMSE ratio: ratio of the root mean squared error (RMSE) between $\widehat Q_{\texttt{CPS}}^{(t),\texttt{np}}$ and $\widehat Q_{\texttt{CPS}}^{(t),\texttt{ipw}}$ (values $<1$ indicate higher efficiency of $\widehat Q_{\texttt{CPS}}^{(t),\texttt{np}}$). Since there is no established asymptotic distribution theory for the IPW estimator, we construct its confidence interval using an \textit{ad hoc} method  (see Appendix~A.4 for details).
    \end{tablenotes}
\end{spacing}
\end{threeparttable}}
\end{table}

\section{Data application: the Add Health study}\label{sec:application}

We apply the proposed nonparametric efficient estimator to the Add Health dataset \citep{harris2019cohort} to evaluate how adolescent smoking influences academic achievement, with  interest in the causal effects across the distribution of GPA. 
Following \cite{qu2021semiparametric}, we construct clusters based on students’ self-reported nominations of close friends. For each participant, we form a cluster of size three ($M_i = 3$), consisting of the student and his or her nominated best female and best male friend. A greedy search algorithm is employed to ensure that no individual appears in more than one cluster, yielding a final analytic sample of 2{,}768 non-overlapping clusters with no missing data. The treatment is defined as the smoking status in the past 12 months ($A_{ij}=1$ if the student smoked; $A_{ij}=0$ otherwise). The outcome $Y_{ij}$ is measured as the student’s total GPA over four core subjects (mathematics, English, history, and science), ranging from 1 to 4 with higher values indicating better academic performance. We adjust for a set of individual-level covariates, including age, sex, race, overall health status, alcohol use, involvement in physical fights within the past year, maternal and paternal education, and maternal and paternal employment status. 

We employ our method to evaluate the CPS policy, $H_{\delta}^{\texttt{CPS}}$, with $\delta\in[0.5,2]$ and outcome quantile $q\in [0.1,0.9]$. The CPS is particularly suitable in this setting where it may be unrealistic to enforce deterministic or universal treatment rules to change adolescent smoking status, but feasible to modify their probability of smoking through programmatic or policy interventions. Note that $\delta<1$ (or $>1$) corresponds to interventions that reduce (or increase) the smoking propensity, and $\delta=1$ represent the factual treatment assignment mechanism. With a slight abuse of notation, for $\delta\in[0.5,2]$, we are interested in studying the overall quantile effect $\text{OQE}_{\texttt{CPS}}(q;\delta,1) = Q_{\texttt{CPS},\delta}^{(\star)}(q) - Q_{\texttt{CPS},\delta=1}^{(\star)}(q)$, direct quantile effect $\text{DQE}_{\texttt{CPS}}(q;\delta) = Q_{\texttt{CPS},\delta}^{(1)}(q) - Q_{\texttt{CPS},\delta}^{(0)}(q)$, and spillover quantile effects $\text{SQE}_{\texttt{CPS},a}(q;\delta,1) = Q_{\texttt{CPS},\delta}^{(a)}(q) - Q_{\texttt{CPS},\delta=1}^{(a)}(q)$, with $a\in\{0,1\}$. We follow the modeling strategies in the simulation study to specify the machine learning models for the nuisance parameters.

\begin{figure}
\begin{center}
\includegraphics[width=0.99\textwidth]{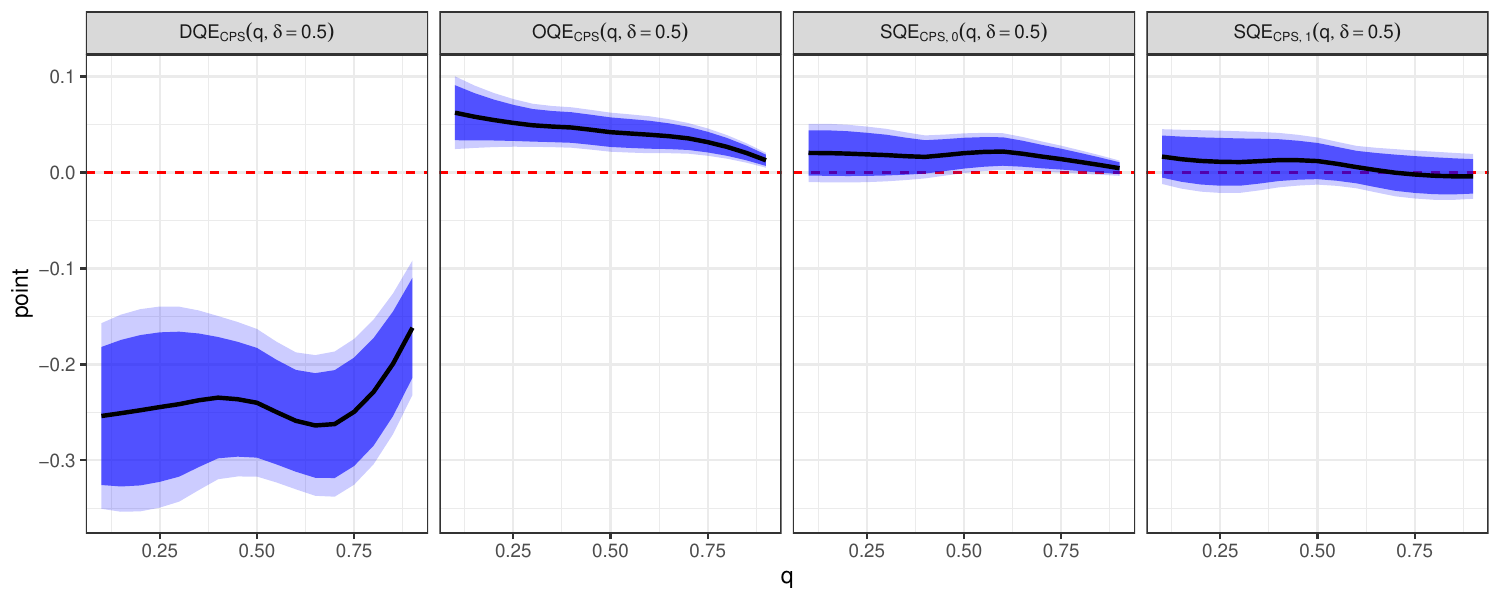}
\end{center}
\caption{The quantile direct effect, overall effect, and spillover effects under the incremental propensity score policy, for $q\in[0.1,0.9]$ with $\delta$ fixed at 0.5. The black line indicates point estimates, the darkblue shaded area indicates the 95\% pointwise confidence interval, and the lightblue shaded area indicates the uniform confidence band over $q\in[0.1,0.9]$.\label{fig:add_health_quantiles} }
\end{figure}

Figure \ref{fig:add_health_quantiles} displays the network quantile causal effects for $q$ ranging from 0.1 to 0.9 under a fixed policy with $\delta=0.5$ when the odds that a cluster contains one more smoker under the policy is halved relative to the factual assignment mechanism. We observe significantly negative direct quantile effects, and the magnitude decreases at higher quantiles, suggesting that a direct change in smoking status has a larger impact on lower GPAs; for example, $\widehat{\text{DQE}}_{\texttt{CPS}}^{\texttt{np}}(0.2;0.5)=-0.247$ (95\% CI: $[-0.326,-0.169]$) compared with $\widehat{\text{DQE}}_{\texttt{CPS}}^{\texttt{np}}(0.9;0.5)=-0.161$ (95\% CI: $[-0.213,-0.109]$). The overall quantile effects comparing the policy ($\delta=0.5$) to the factual regime ($\delta=1$) are positive and significant across all quantiles considered, but again the effect size diminishes for higher quantiles. For example, reducing the cluster propensity by half (on the odds ratio scale) yields a GPA increase of about 0.062 points at $q=0.1$ ($\widehat{\text{OQE}}_{\texttt{CPS}}^{\texttt{np}}(0.1;0.5)=0.062$) but only about 0.013 points at $q=0.9$ ($\widehat{\text{OQE}}_{\texttt{CPS}}^{\texttt{np}}(0.9;0.5)=0.013$). We also observe positive spillover effects when conditioning on the student being a non-smoker (i.e., $\text{SQE}_{\texttt{CPS},0}$). For instance, $\widehat{\text{SQE}}_{\texttt{CPS},0}^{\texttt{np}}(0.5;0.5)=0.020$ (95\% CI: $[0.004,0.036]$) indicates that, when a student does not smoke, reducing friends' smoking propensity by half (on the odds ratio scale) increases the median GPA by 0.020 points. By contrast, the spillover effect conditioning on the student being a smoker (i.e., $\text{SQE}_{\texttt{CPS},1}$) is statistically insignificant. Thus, if the student himself is a smoker, changing his friends' smoking has minimal impact for his academic performance.

\begin{figure}
\begin{center}
\includegraphics[width=0.99\textwidth]{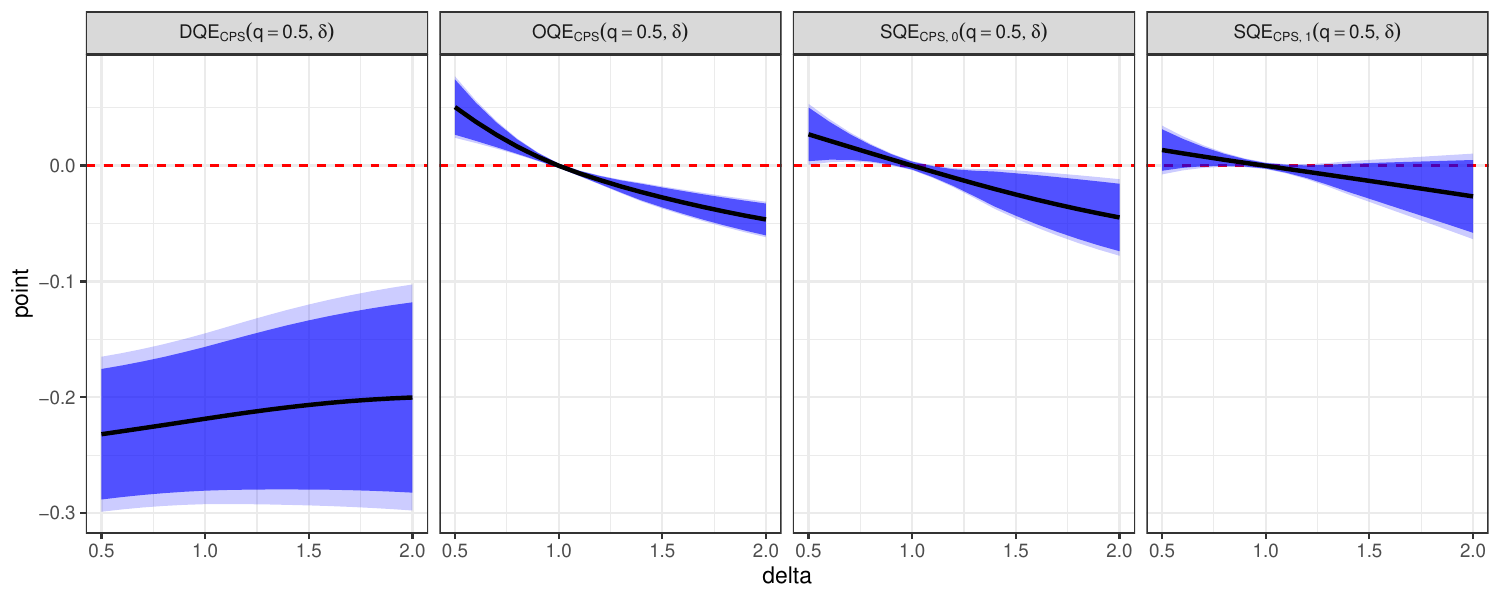}
\end{center}
\caption{The quantile direct effect, overall effect, and spillover effects under the incremental propensity score policy, for $\delta\in[0.5,2.0]$ with $q$ fixed at 0.5. The black line indicates point estimates, the darkblue shaded area indicates the 95\% pointwise confidence interval, and the lightblue shaded area indicates the uniform confidence band over $\delta\in[0.5,2.0]$.\label{fig:add_health_policies} }
\end{figure}


Figure \ref{fig:add_health_policies} displays the median network causal effects for $\delta$ ranging from 0.5 to 2.0 at a fixed quantile level of $q=0.5$. The median direct effect is consistently negative and statistically significant, indicating that smoking always lowers GPA. We also observe a negative overall quantile effect for $\delta<1$ and a positive overall quantile effect for $\delta>1$, with confidence intervals excluding zero, suggesting that relative to the factual assignment mechanism under which the cluster maintains its observed smoking propensity, interventions that decrease (increase) smoking propensity improve (impair) median GPA. For the spillover effect conditional on the student being a non-smoker (i.e., $\text{SQE}_{\texttt{CPS},0}$), we similarly find a positive and significant effect when $\delta<1$ and a negative and significant effect when $\delta>1$, with larger magnitudes at more extreme values of $\delta$; for example, when the student does not smoke, halving or doubling the cluster propensity (on the odds ratio scale) increases the student's median GPA by 0.026 points (95\% CI: $[0.003,0.050]$) or decreases it by 0.045 points (95\% CI: $[-0.073,-0.015]$), respectively. In contrast, the spillover effect conditional on the student being a smoker (i.e., $\text{SQE}_{\texttt{CPS},1}$) is not statistically significant. Similar pattern of results extends to other quantile levels (see Supplementary Material Figures 1--2 for network quantile causal effect with $q=0.2$ and 0.8, respectively).

\section{Discussion}\label{sec:discussion}


The Add Health analysis reveals insights that a focus on average effects alone would obscure. The direct quantile effects of smoking on GPA are substantially larger at lower quantiles, suggesting that smoking disproportionately harms academically vulnerable students who may lack protective factors that buffer higher-performing peers. This distributional heterogeneity would be entirely masked by a single average causal effect, potentially leading to suboptimal policy targeting. The asymmetry between spillover effects also bears important implications. Among non-smokers, reducing peers' smoking propensity significantly improves GPA, whereas among smokers, peer smoking has negligible additional impact, likely because the student's own smoking already dominates the effect on academic performance. This suggests that school-based anti-smoking interventions targeting peer networks may be most effective for improving outcomes among non-smoking students. More broadly, when outcomes are skewed or heavy-tailed, as is common with academic performance, earnings, or health expenditure data, network quantile causal effects provide a more complete and policy-relevant characterization than existing methods in the interference literature, which have focused almost exclusively on mean potential outcomes.

It is worth noting that our nonparametric identification relies on the treatment ignorability assumption, which may be violated in observational studies due to unmeasured confounders. While sensitivity methods exist for quantile causal effects \citep{cheng2024doubly}, we are unaware of analagous framework for quantile causal effects with partial interference and this merits future research. Second, the partial interference assumption restricts spillover to within predefined clusters, which may not address applications with diffuse network connections. Extending the proposed framework to accommodate general network interference structures, for example, through exposure mapping strategies, would be of interest and will be reported in future work.

\section*{Acknowledgement}
Research in this article was supported by the Patient-Centered Outcomes Research Institute\textsuperscript{\textregistered} (PCORI\textsuperscript{\textregistered} Award ME-2023C1-31350). The statements presented in this article are solely the responsibility of the authors and do not necessarily represent the views of PCORI\textsuperscript{\textregistered}, its Board of Governors or Methodology Committee. 

\section*{Supplementary material}
Supplementary Material includes appendices and  supplementary figures and tables, which is available at \url{https://www.chaochengstat.com}.

\spacingset{1}

\bibliographystyle{jasa3}

\bibliography{Bibliography-MM-MC}

\end{document}